\documentclass{aastex}          
\usepackage{spr-astr-addons}    
\usepackage{natbib}             
\usepackage{multirow}
\usepackage{bigstrut}
\usepackage{graphicx}
\usepackage{eqnarray,amsmath}
\usepackage{amsmath}
\usepackage{ulem}

\begin{document}

\title{Star-formation rate in compact star-forming galaxies}

\shortauthors{I. Y. Izotova, Y. I. Izotov}

\shorttitle{
SFR in compact star-forming galaxies
}


\author{I. Y. Izotova}
\affil{Astronomical Observatory of Taras Shevchenko Kyiv National University\\
Observatorna str., 3, 04053, Kyiv, Ukraine\\
tel: +380444860021, fax: +380444862191\\ e-mail:izotova@observ.univ.kiev.ua}
\email {izotova@observ.univ.kiev.ua}

\author{Y. I. Izotov}
\affil{Main Astronomical Observatory of the National Academy of Sciences of
Ukraine\\
Zabolotnoho str., 27, 03143, Kyiv, Ukraine\\
tel: +380445264771, fax: +380445262197\\ e-mail:izotov@mao.kiev.ua}
\email {izotov@mao.kiev.ua}

\begin{abstract}
We use the data for the H$\beta$ emission-line, far-ultraviolet (FUV) 
and mid-infrared 22 $\mu$m continuum luminosities to estimate star 
formation rates $<$SFR$>$ averaged over the galaxy lifetime for a sample of 
about 14000 bursting compact star-forming galaxies (CSFGs) selected from the 
Data Release 12 (DR12) of the Sloan Digital Sky Survey (SDSS).
The average coefficient linking $<$SFR$>$ and the star formation rate SFR$_0$ 
derived from the H$\beta$ luminosity at zero starburst age is found 
to be 0.04. We compare $<$SFR$>$s with some commonly used SFRs which 
are derived adopting a continuous star formation during a period of 
$\sim$ 100 Myr, and find that the latter ones are 2 -- 3 times higher.
It is shown that the relations between
SFRs derived using a geometric mean of two
star-formation indicators in the UV and IR ranges and 
reduced to zero starburst age have
considerably lower dispersion compared to those with single 
star-formation indicators. We suggest that our
relations for $<$SFR$>$ determination are more appropriate for CSFGs because
they take into account a proper temporal evolution of their luminosities.
On the other hand, we show that commonly used SFR relations can be 
applied for approximate estimation within a factor of $\sim$ 2 of the $<$SFR$>$ 
averaged over the lifetime of the bursting compact galaxy.
\end{abstract}


\section{Introduction}\label{s:Introduction}

The star formation rate (SFR) is one of the most important parameters
regulating the galaxy evolution. Many studies have been made in the past to
derive SFRs from the luminosities at various 
wavelengths, spanning from the far-UV (FUV), where the massive stars emit the 
bulk of their energy, to the infrared, where the dust-reprocessed light from 
those stars emerges, and to the radio, which is a tracer of supernova activity 
at wavelengths $\lambda$$\ga$20 cm, while free-free emission from H~{\sc ii}
regions dominates at wavelengths shorter than 6 cm.

\begin{figure*}[t]
\hbox{
\includegraphics[angle=-90,width=0.49\linewidth]{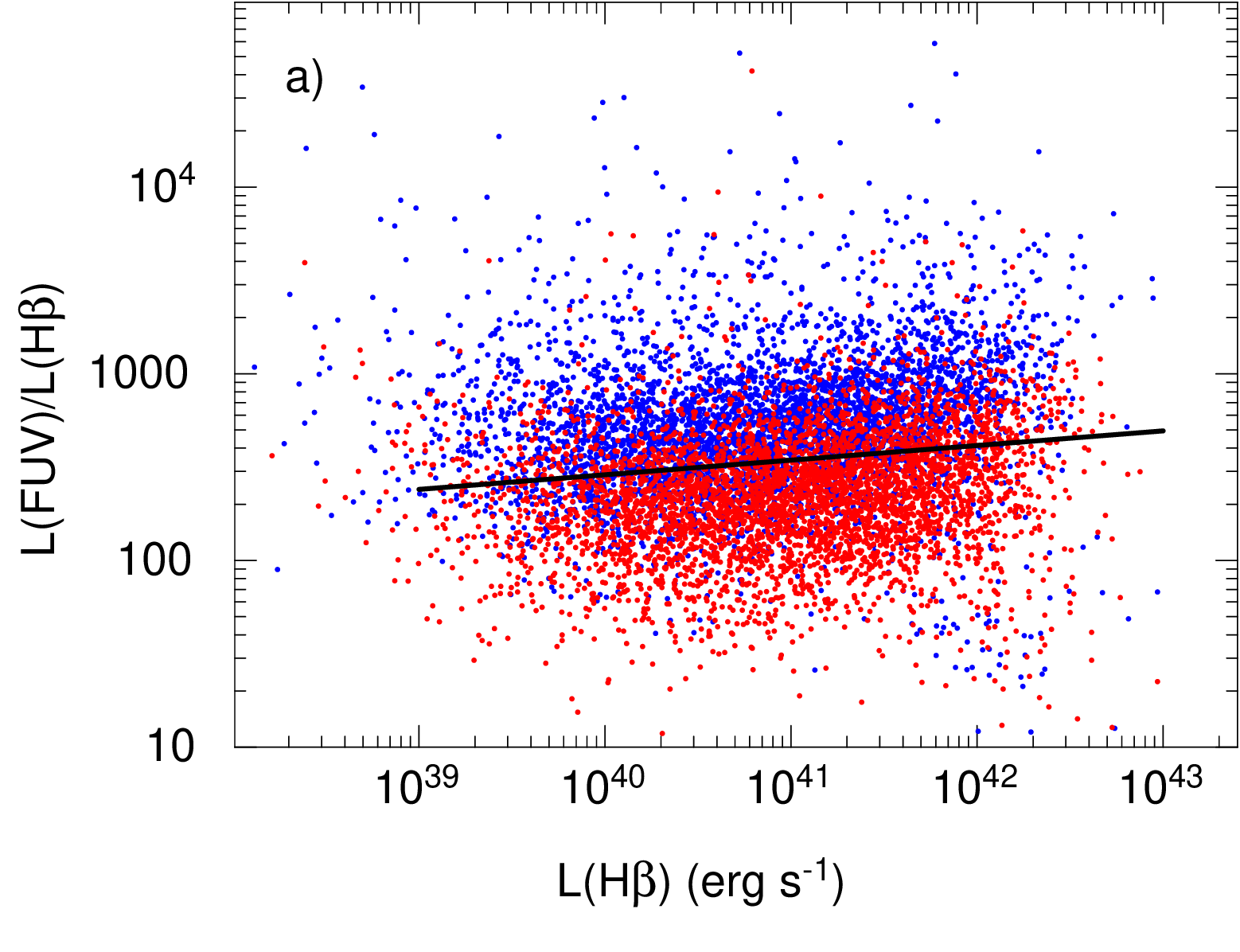}
\includegraphics[angle=-90,width=0.49\linewidth]{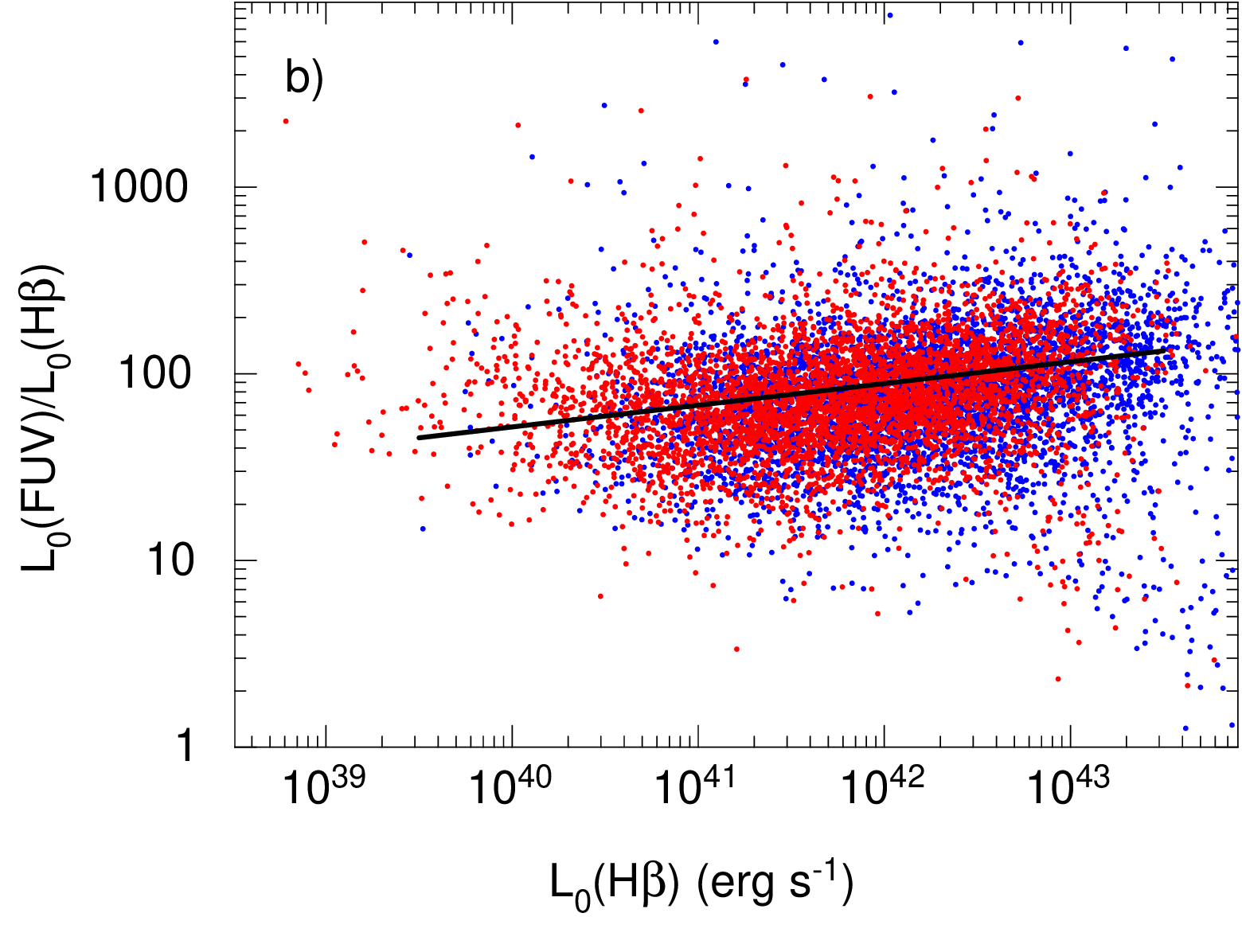}
}
\caption{{\bf a)} The relation between the
FUV-to-H$\beta$ luminosity ratio and the H$\beta$ luminosity.
CSFGs with EW(H$\beta$) $\geq$ 50\AA\ and EW(H$\beta$) $<$ 50\AA\ are shown
by 
red and blue
symbols, respectively. The solid line is the
maximum-likelihood relation log[$\nu L_\nu $(FUV)/$L$(H$\beta$)] = 
0.078$\times$log$L$(H$\beta$) $-$ 0.68 for the entire sample. {\bf b)}
Same as in {\bf a)}, but luminosities are reduced to a zero starburst age. The 
solid line is the maximum-likelihood relation 
log[$\nu L_{\nu,0} $(FUV)/$L_0$(H$\beta$)] = 
0.12$\times$log$L_0$(H$\beta$) $-$ 2.94 for the entire sample. H$\beta$
luminosities in both panels are expressed in erg s$^{-1}$}
\label{fig1}
\end{figure*}

\begin{figure}
\includegraphics[angle=-90,width=0.99\linewidth]{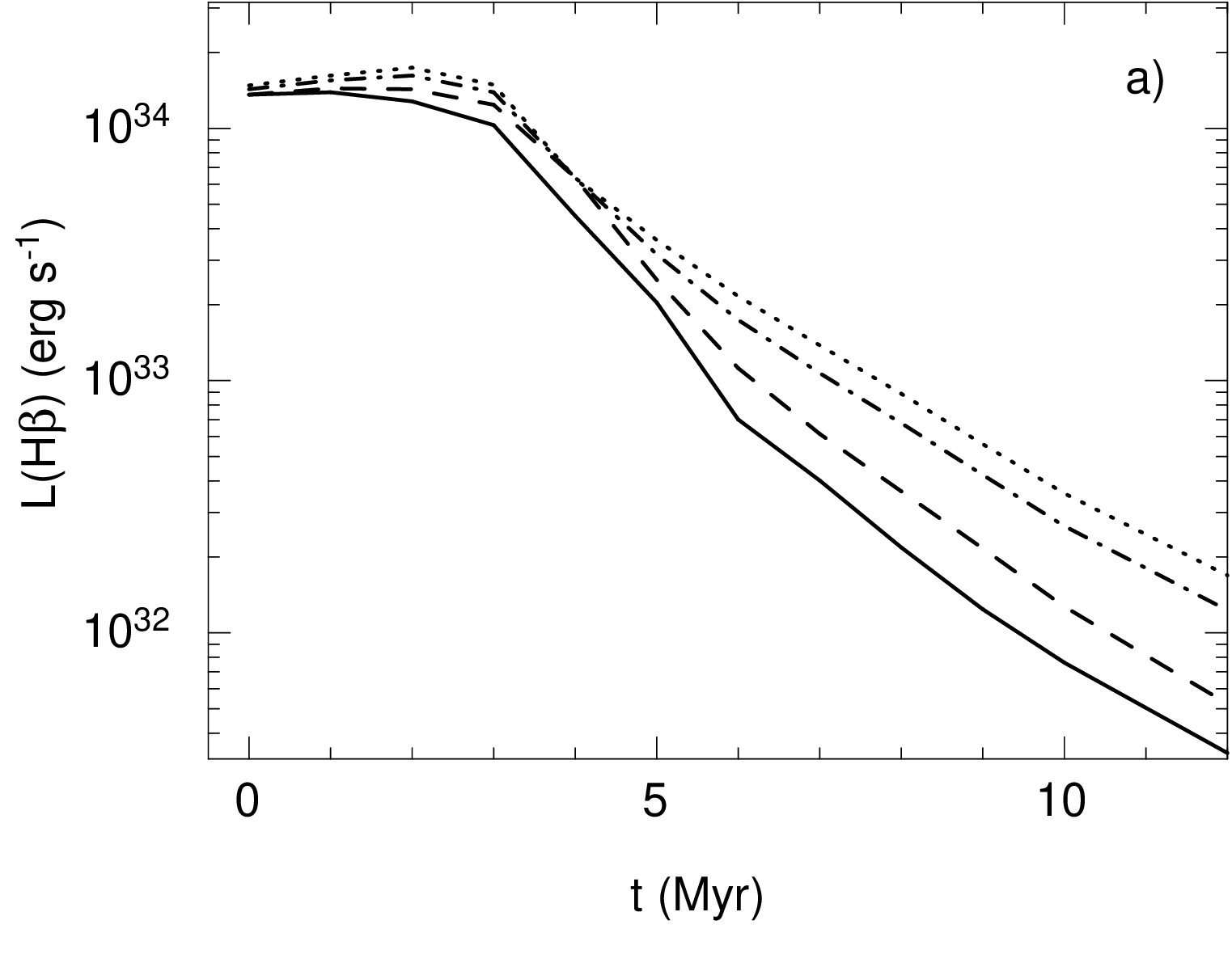}
\includegraphics[angle=-90,width=0.99\linewidth]{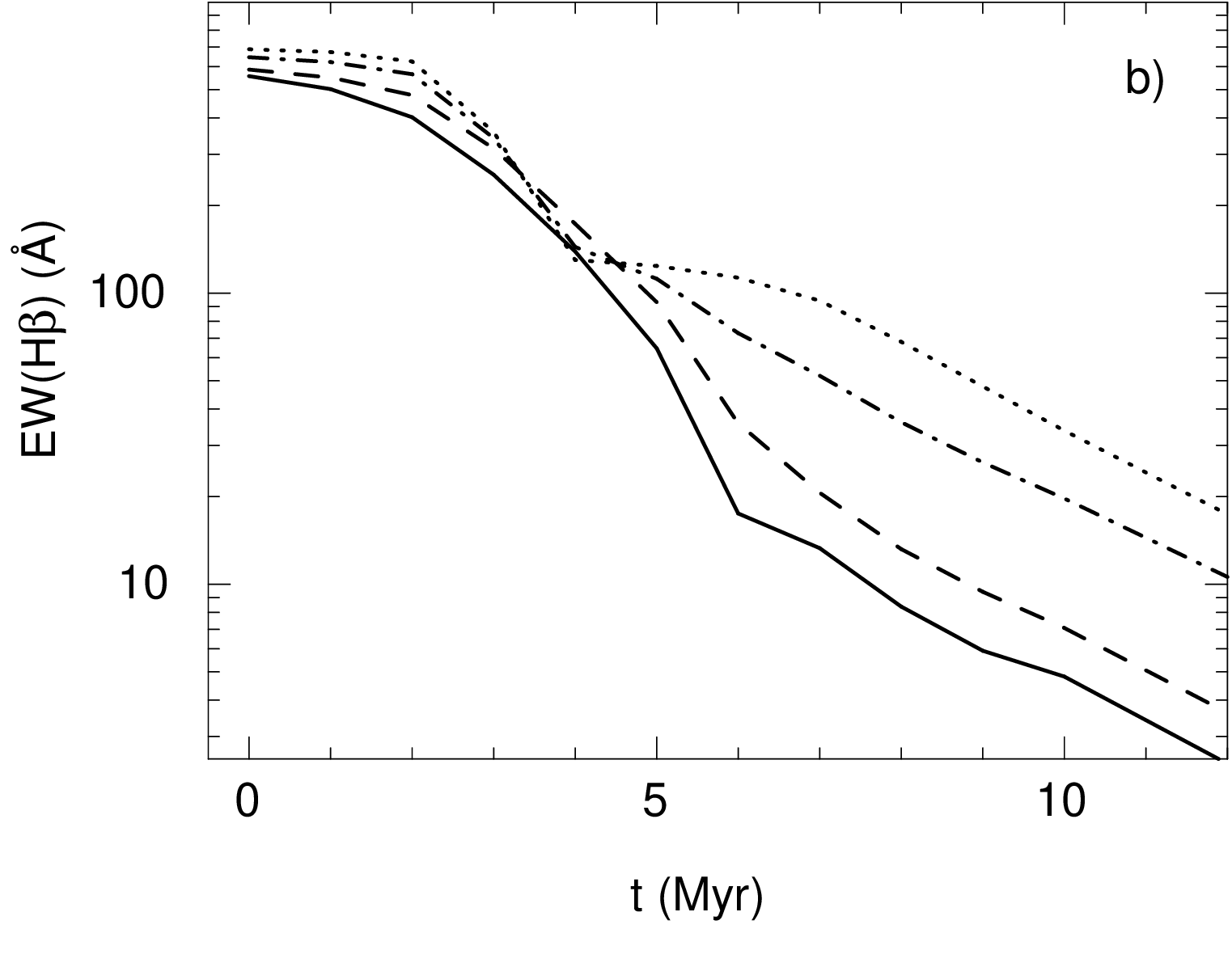}
\includegraphics[angle=-90,width=0.99\linewidth]{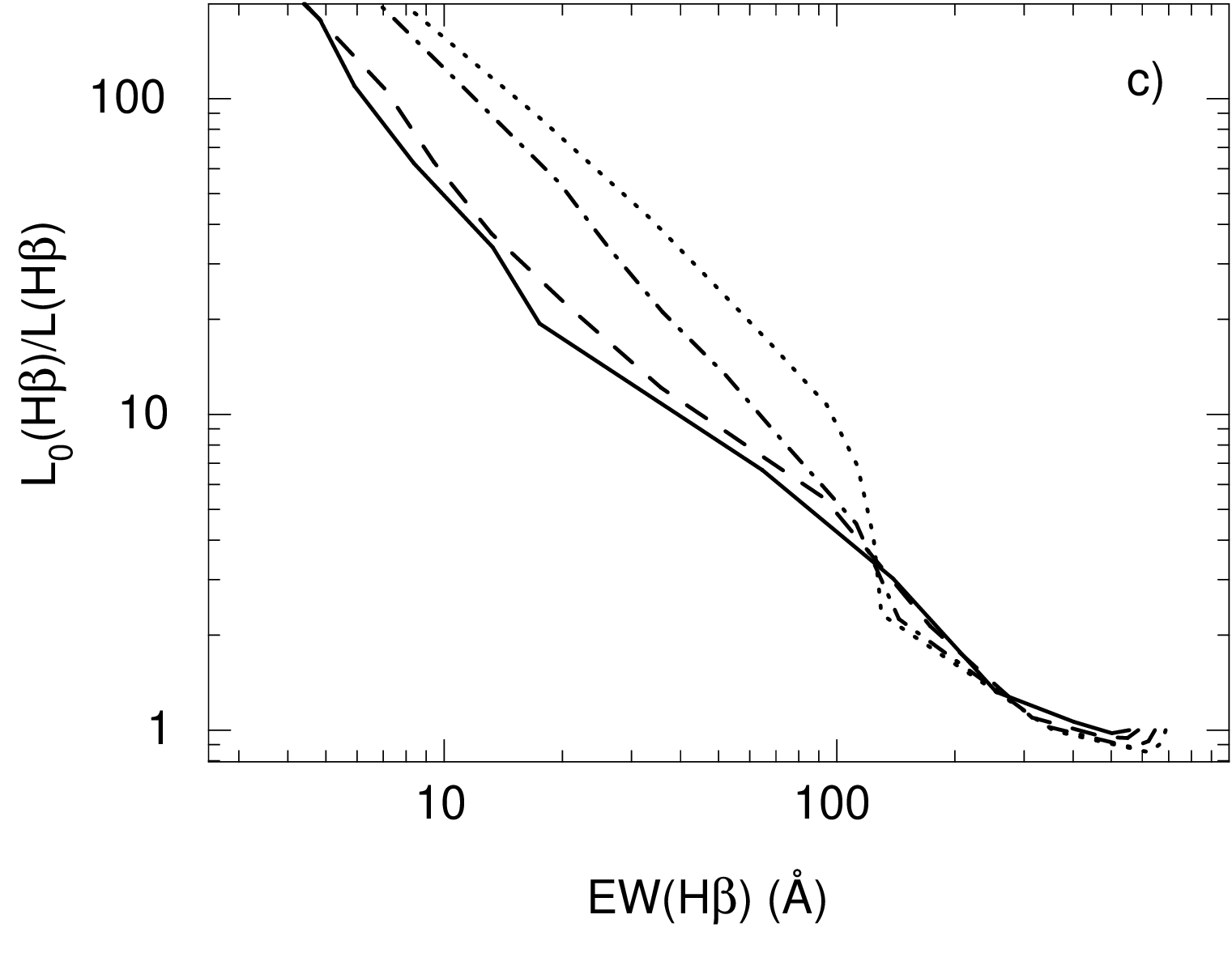}
\caption{{\bf a)} The dependences of the H$\beta$ luminosity on the starburst
age for instantaneous burst models with various metallicities: 1/50 solar
(dotted line), 1/20 solar (dash-dotted line), 1/5 (dashed line), and 1/2.5 solar
(solid line) \citep{L99}. Luminosities are derived for a starburst with 
a mass of 1 M$_\odot$. {\bf b)} Same as in {\bf a)} but for the relation
between the H$\beta$ equivalent width EW(H$\beta$) and the starburst age.
{\bf c)} The dependences of the H$\beta$ luminosity correction factor on
the H$\beta$ equivalent width}
\label{fig2}
\end{figure}

Various luminosities, which serve as SFR monochromatic 
indicators, have been considered by e.g. \citet{K98}
and \citet{C10}, while combinations of the luminosities at two or more 
different wavelengths 
have been analysed in other studies \citep[e.g. ][]{K09,H11,L13,B16,P16}. 
The luminosities of 
recombination H$\alpha$, H$\beta$ and other hydrogen lines, UV, 
mid-infrared (MIR), far-infrared (FIR) and radio
continuum luminosities are among the most commonly used indicators. 
Emission at these various wavelengths is generated by
stars with different lifetimes, with the shortest lifetimes for the most massive
stars producing hydrogen recombination lines.
These SFR indicators have been applied to different samples of SF galaxies at 
redshifts $z$ from 0 to $\sim$6 
\citep{R10,R15,C13,J13,K13,B15,P15,S15,S16,D17,K17} and for samples with
various physical properties.

SFR calibrations are affected by many input factors 
\citep[discussed e.g. by ][]{Z13}.
The initial mass function (IMF), metallicity, binary 
interactions, evolution models of rotating and non-rotating stars and stellar
atmosphere models are among them. Some of these factors are not well known and 
various assumptions on them may lead to SFR variations of $\sim$ 0.3 dex.
One of the important factors influencing SFR is the star formation history.

The SFR calibrations are usually obtained from population synthesis models
adopting an IMF and prolonged star-formation activity. 
In particular, several widely used calibrations \citep[e.g. ][]{K98,C10} are 
obtained assuming the \citet{S55} IMF and relatively large period of 
$\sim$ 100 Myr for star formation with a 
constant rate. With this relatively long period, the luminosities in different 
wavelength ranges and their ratios are slowly changing with time. Besides that
some other SF histories, e.g. with exponentially decaying or rising SFRs, were 
also considered in some papers. 

However, such an assumption may not be valid in the case of compact star-forming
galaxies (CSFGs) with star-formation proceeded in short starbursts, which 
dominate in the galaxy light, but quickly decay on the time scales of $\la$ 10 
Myr. Indeed, Starburst99 models \citep{L99,L14} for instantaneous bursts 
predict that the H$\beta$ luminosity decreases by two orders of 
magnitude during the first 10 Myr. Therefore, the definition of the SFR for 
starburst galaxies is somewhat uncertain because of the uncertainties in the 
determination of the period of star formation. Because of this the use of the 
SFR calibrations discussed above for starburst galaxies and the 
comparison of their SFRs with those for
star-forming galaxies with prolonged star formation are questionable.

In this paper we propose a recipe for the determination of the SFR averaged
over the lifetime of the bursting compact galaxies and discuss the 
applicability of commonly used
SFR calibrations for CSFGs with strong emission lines where star-formation is 
occurred in short intense starbursts. \citet{I16a} have constructed this CSFGs 
sample from the Sloan Digital Sky Survey (SDSS). Selection
criteria for the sample and its general characteristics are considered
in Sect.~\ref{s:Sample}. We discuss the star formation history in CSFGs in 
Sect.~\ref{SFH}. The relations for the SFR determination at a zero starburst 
age are considered in Sect.~\ref{SFR0}. In Sect.~\ref{SFRcomp} the derived 
SFRs averaged over the CSFG lifetime are compared with commonly used SFR 
calibrations. We summarize our results in Sect.~\ref{concl}.

\section{The sample}\label{s:Sample}

\citet{I16a} have constructed a sample of $\sim$14000 CSFGs from the 
spectroscopic data base of the SDSS Data Release 12 (DR12) \citep{A15}
which includes classic SDSS subsample from Data Releases 9 and earlier
releases and BOSS subsample from Data Releases 10 and 12. The spectra of these
two different subsamples were obtained with two different round spectroscopic
apertures, 3 arcsec and 2 arcsec in diameter, respectively. Furthermore,
spectra from the classic SDSS and BOSS subsamples cover different wavelength
ranges, $\sim$ 3800 -- 9200\AA\ and $\sim$ 3600 -- 10000\AA, respectively.

The objects were selected by their compactness adopting the Petrosian
radius $R_{50}$ $\la$ 3 arcsec to minimize aperture corrections
in the comparison of spectroscopic and photometric data at various 
wavelengths from UV to mid-infrared, by strong emission lines as defined by
the H$\beta$ equivalent width EW(H$\beta$) $\ga$ 10\AA, and by absence of AGN 
spectral features. CSFGs are distributed over the redshift range $z$~=~0~--~1
with a median value of 0.2. The median value of EW(H$\beta$) for the sample is 
$\sim$40\AA, while a considerable fraction of CSFGs ($\sim$10 percent) 
is characterized by much higher EW(H$\beta$) $>$ 100\AA.
We note that strong
H$\delta$, H$\gamma$ and H$\beta$ emission lines are seen in spectra
of all galaxies from our sample, regardless of the redshift. As for 
H$\alpha$, it is present in spectra
of galaxies with $z <$ 0.4 from the classic SDSS subsample and with 
$z$ $<$ 0.52 from the BOSS subsample. The presence of several hydrogen lines
allows the reliable simultaneous determination of extinction and underlying
stellar absorption from the observed Balmer decrement, following to 
\citet{I94}. Correspondingly, fluxes of hydrogen lines were corrected for
both effects. In general, correction for underlying absorption in CSFGs
is small, not exceeding $\sim$ 10 percent on average 
for the H$\beta$ emission line.

We use the method described by \citet{I16a} to derive stellar masses
of CSFGs from SED fitting of the SDSS spectra. The method includes
both the stellar and nebular continua and is outlined below. 
The star-formation history is approximated 
assuming a short burst with age $t_{\rm young}$ $<$ 10 Myr and a continuous star 
formation with a constant SFR for the older population with a constant SFR 
during the time interval between $t_i$ and $t_f$ ($t_f$ $<$ $t_i$ and zero age 
is now). Thus, the age of the oldest stars in the model is $t_{\rm old}$ $\equiv$ 
$t_i$. The contribution of each stellar population to the SED
was parameterized by the ratio $b$ = $M_{\rm young}$/$M_{\rm old}$, where $M_{\rm young}$
and $M_{\rm old}$ are the masses of the young and old stellar populations
formed during the recent burst with $t_{\rm young}$ $<$ 10 Myr and the prior 
continuous star formation between $t_f$ and $t_i$, respectively.


We note that star formation rate in CSFGs is uncertain at large ages and
may not be constant. \citet{W11} analyzed resolved stellar populations of
dwarf galaxies of various morphological types in the Local Volume
and concluded that the star formation histories (SFHs) 
are complex and the mean values are inconsistent with simple SFH models, 
e.g., single bursts, constant SFRs, or smooth, exponentially declining SFRs.
However, results by \citet{W11} are obtained for low-mass and relatively 
quiescent galaxies, while our CSFGs are more massive and experience vigorous 
recent SF episode.
Furthermore, emission in the optical SDSS spectra used for SED fitting
is strongly dominated by the young stellar population. This makes the
stellar mass determination of the old stellar population even more uncertain.
In spite of these uncertainties we adopted the simplest SFH with a constant
SFR for the old stellar population and rely mainly on the statistical 
properties of the sample.

To fit the SED we carried out a series of 5000 Monte Carlo simulations for 
each galaxy by randomly varying $t_{\rm y}$, $t_i$, $t_f$, and $b$.  
A  grid  of  instantaneous  burst  SEDs  with  a
wide  range  of  ages  from  0  Myr  to  15  Gyr  and various metallicities
was  calculated with Starburst99 \citep{L99,L14} to  derive
the SED of the galaxy stellar component. Input parameters
for  the  grid  calculations  included  Padova  stellar  evolution
tracks \citep{G00}, models of stellar atmospheres
by \citet{L97} and \citet{S92}, and the stellar initial mass function of
\citet{S55} with the upper and low stellar mass limits of 100 M$_\odot$
and 0.1 M$_\odot$, respectively. Then the stellar SED
with any star-formation history can be obtained by integrating the 
instantaneous burst SEDs over time with a specified
time-varying SFR. 

Given the electron temperature
$T_{\rm e}$ in the H~{\sc ii} region, we interpolated emissivities by 
\citet{A84} for the nebular continuum in the Te range of 5000 -- 20000 K.
Then the nebular continuum luminosity at any wavelength
is derived from the observed H$\beta$ luminosity. We also calculate the
equivalent  width  EW(H$\beta$) for  each  model,
integrating over the time luminosities of emission lines and
adjacent  continua.  

The  best  solution  is  required  to  fulfill
the following conditions. First, only models, in which
EW(H$\beta$) is in agreement with the observed value  within  10  percent, 
were  selected.  Second,  the  best
modelled SED satisfying first condition was found from $\chi^2$
minimization of the deviation between the modelled and
the observed continuum.

Comparing different sets of Monte Carlo realizations for the same galaxy
we find that the dispersion of the derived stellar masses is less than 0.1 dex.
In our calculations we do not take into account the stellar mass returned to
the interstellar medium via mass loss. It is estimated to be less than
10 -- 15\%, resulting in the lower stellar mass by the same amount. However, 
we neglect this small reduction of the stellar mass because of uncertainties 
in its determination.

The distribution of CSFG stellar masses is peaked
at $M_\star$ $\sim$ 10$^9$M$_\odot$, while the distribution of H$\beta$
luminosities is peaked at $L$(H$\beta$) $\sim$ 10$^{41}$ erg s$^{-1}$, 
corresponding to SFR $\sim$ 1 M$_\odot$ yr$^{-1}$ as derived from the 
\citet{K98} relation. Thus, on average, the time of only
$\sim$ 1 Gyr is needed to
build up the mass of the CSFG with the SFR derived from the H$\beta$ luminosity.
This is an indication of the bursting nature of star formation in these
galaxies. Furthermore, since the H$\beta$ luminosity in CSFGs is nearly 
proportional to their stellar mass, this conclusion holds for CSFG with any
stellar mass. On the other hand, according to SED fitting, some fraction of 
stars with ages older than 1~Gyr is present in most CSFGs.
Therefore, an average SFR in bursting galaxy may be lower than that 
derived from the observed H$\beta$ luminosity with the \citet{K98} 
relation.

The SDSS spectroscopic data are supplemented by the photometric data 
obtained with the {\sl Galaxy Evolution Explorer} ({\sl GALEX}) in the UV 
range and the {\sl Wide-field Infrared Survey Explorer} ({\sl WISE}) in the 
MIR range. We note that nearly 70 percent of CSFGs were
detected by both the {\sl GALEX} and {\sl WISE}. The SDSS spectra are used 
in this paper to derive galaxy stellar masses $M_\star$, H$\beta$ luminosities 
$L$(H$\beta$), star formation rates SFR(H$\beta$), and equivalent widths 
EW(H$\beta$), while {\sl GALEX} and {\sl WISE} data are used for the 
determination of FUV and 22$\mu$m luminosities, SFR(FUV) and SFR(22$\mu$m).

The luminosities are derived adopting distance 
obtained from the NED cosmological
calculator \citep{W06} for the cosmological parameters $H_0$ 
= 67.1 km s$^{-1}$ Mpc$^{-1}$, $\Omega_m$ = 0.318, $\Omega_\Lambda$ = 0.682
\citep{P14} and a flat geometry. The correction of FUV and H$\beta$ 
luminosities for extinction was done it two steps. Firstly, they were corrected
for the Milky Way extinction at the observed wavelengths adopting 
$A(V)_{\rm MW}$ from the NED. Secondly, correction for the internal galaxy
extinction derived from the hydrogen Balmer decrement in the SDSS spectra 
after its correction for the Milky Way extinction was applied at the
rest-frame wavelengths. In all cases a reddening law by \citet{C89} and a 
total-to-selective extinction ratio $R(V)$~=~3.1 were adopted. 
Finally, H$\beta$ luminosities
were corrected for the spectroscopic aperture by using the correction factor
of 2.512$^{r({\rm ap})-r}$, where $r$ and $r$(ap) are the total SDSS magnitude and
the magnitude within the spectroscopic aperture of 3 arcsec for classic 
SDSS objects and of 2 arcsec for BOSS objects.

\section{Star formation history in CSFGs}\label{SFH}   

The comparison of the ultraviolet-to-H$\beta$ luminosity ratios in
galaxies with different equivalent widths EW(H$\beta$) of the H$\beta$ emission 
line has demonstrated that star formation in CSFGs is bursting in nature 
\citep{I16a}. To illustrate this we show in Fig. \ref{fig1}a
the dependence of the FUV-to-H$\beta$ 
luminosity ratio on the H$\beta$ luminosity 
for CSFGs. It is seen 
that there is an offset between CSFGs with high and low EW(H$\beta$)s 
(red 
and blue symbols, 
respectively). \citet{I16a} assumed that this difference
is due to that the star formation in CSFGs occurred in short starbursts. This
assumption is further supported by the fact that UV emission in these
galaxies is mainly produced in one bright compact SF region as it is
evidenced by the
near-UV (NUV) acquisition images of some CSFGs obtained with the Cosmic 
Origins Spectrograph 
(COS) onboard of the {\sl Hubble Space Telescope} ({\sl HST}) 
\citep{I16b,I18,Y17}.

\begin{figure}[t]
\includegraphics[angle=-90,width=0.98\linewidth]{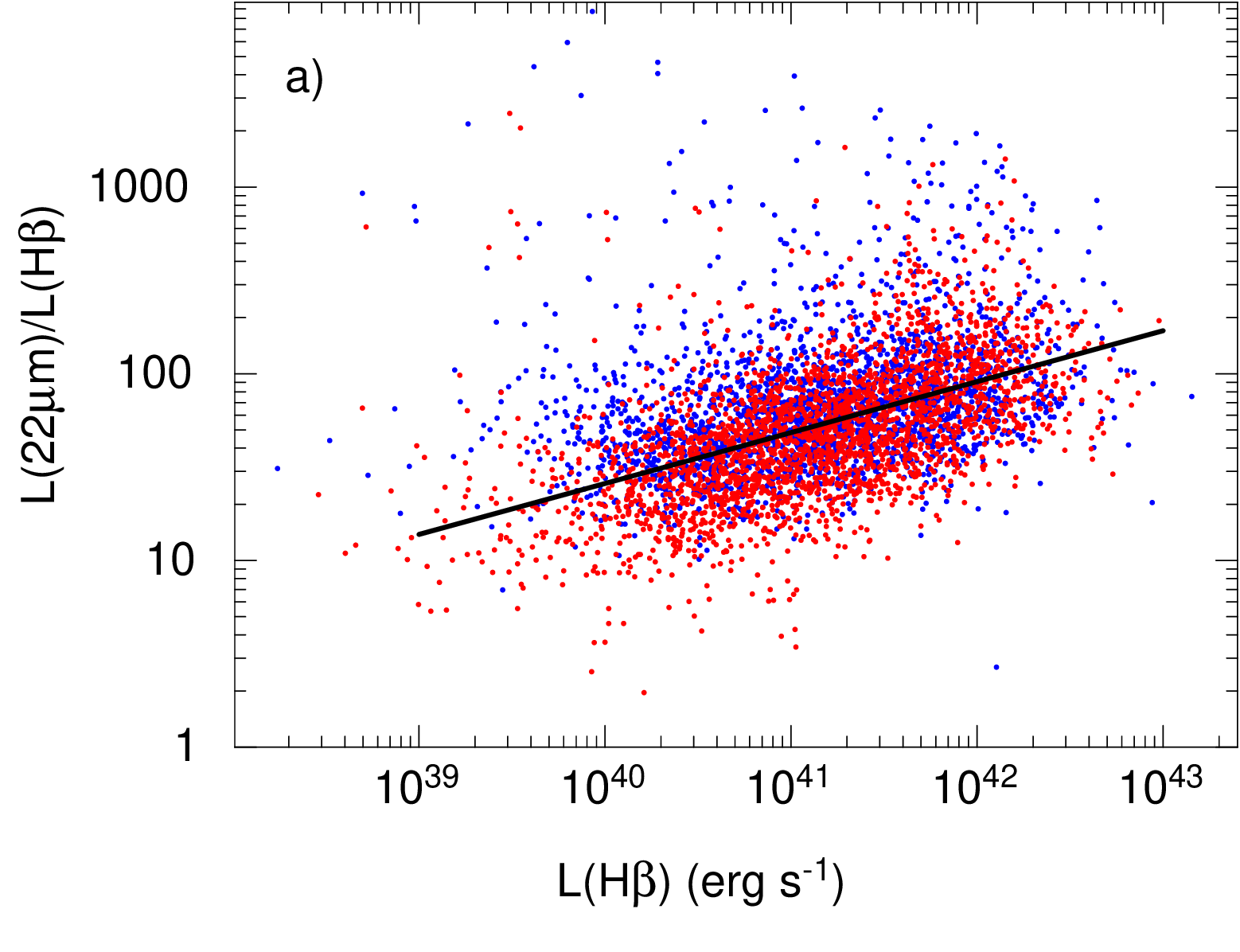}
\includegraphics[angle=-90,width=0.98\linewidth]{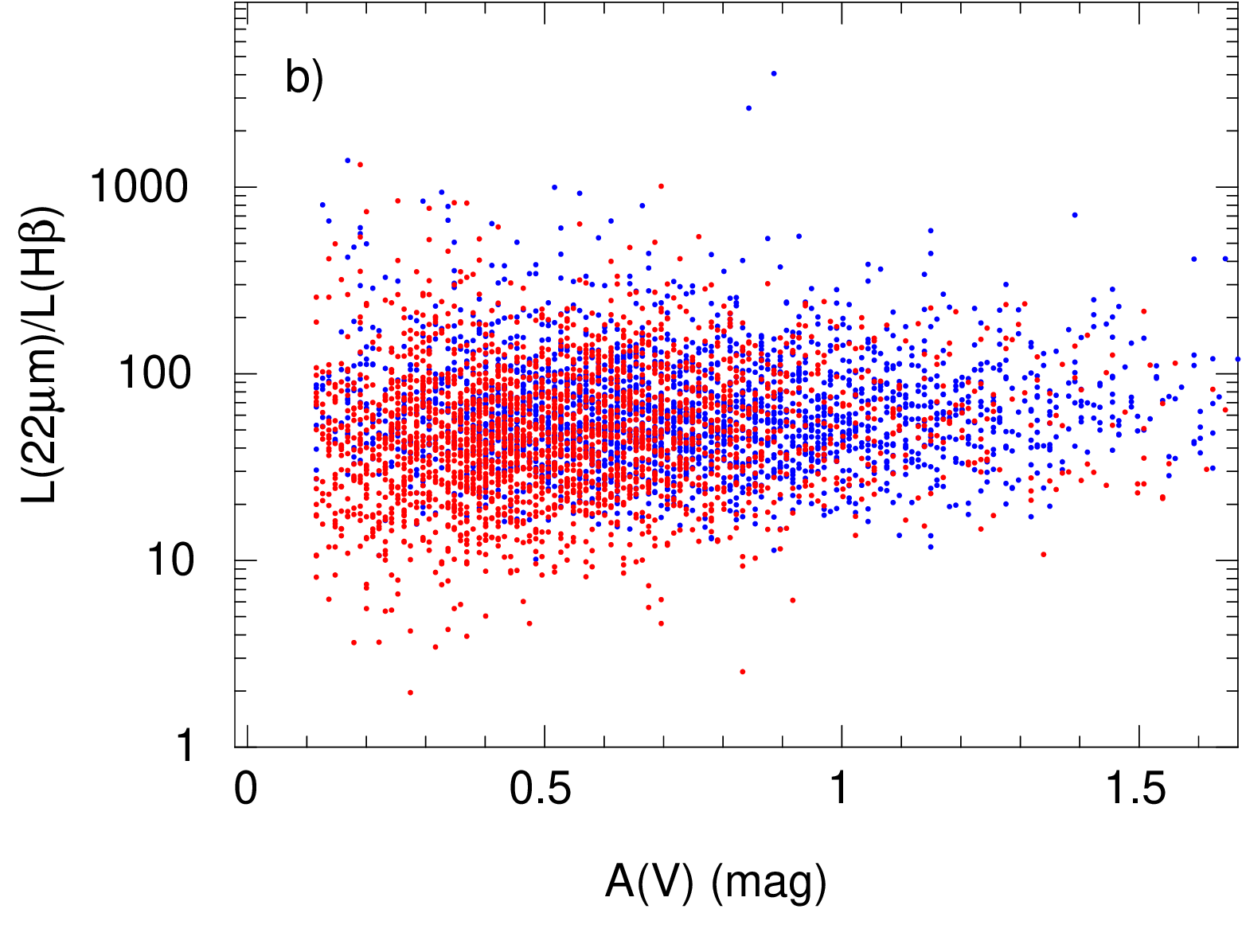}
\caption{{\bf a)} The relation between the 
22$\mu$m-to-H$\beta$ luminosity ratio and the H$\beta$ 
luminosity. CSFGs with EW(H$\beta$) $\geq$ 50\AA\ and EW(H$\beta$) $<$ 50\AA\ 
are shown by 
red and blue 
symbols, respectively. The solid line is 
the maximum-likelihood relation log[$\nu L_\nu $(22$\mu$m)/$L$(H$\beta$)] = 
0.35$\times$log$L$(H$\beta$) $-$ 9.50 for the entire sample. The H$\beta$
luminosities are expressed in erg s$^{-1}$. {\bf b)} The dependence of the
22$\mu$m-to-H$\beta$ luminosity ratio on the extinction $A(V)$}
\label{fig3}
\end{figure}

\begin{figure}[t]
\includegraphics[angle=-90,width=0.98\linewidth]{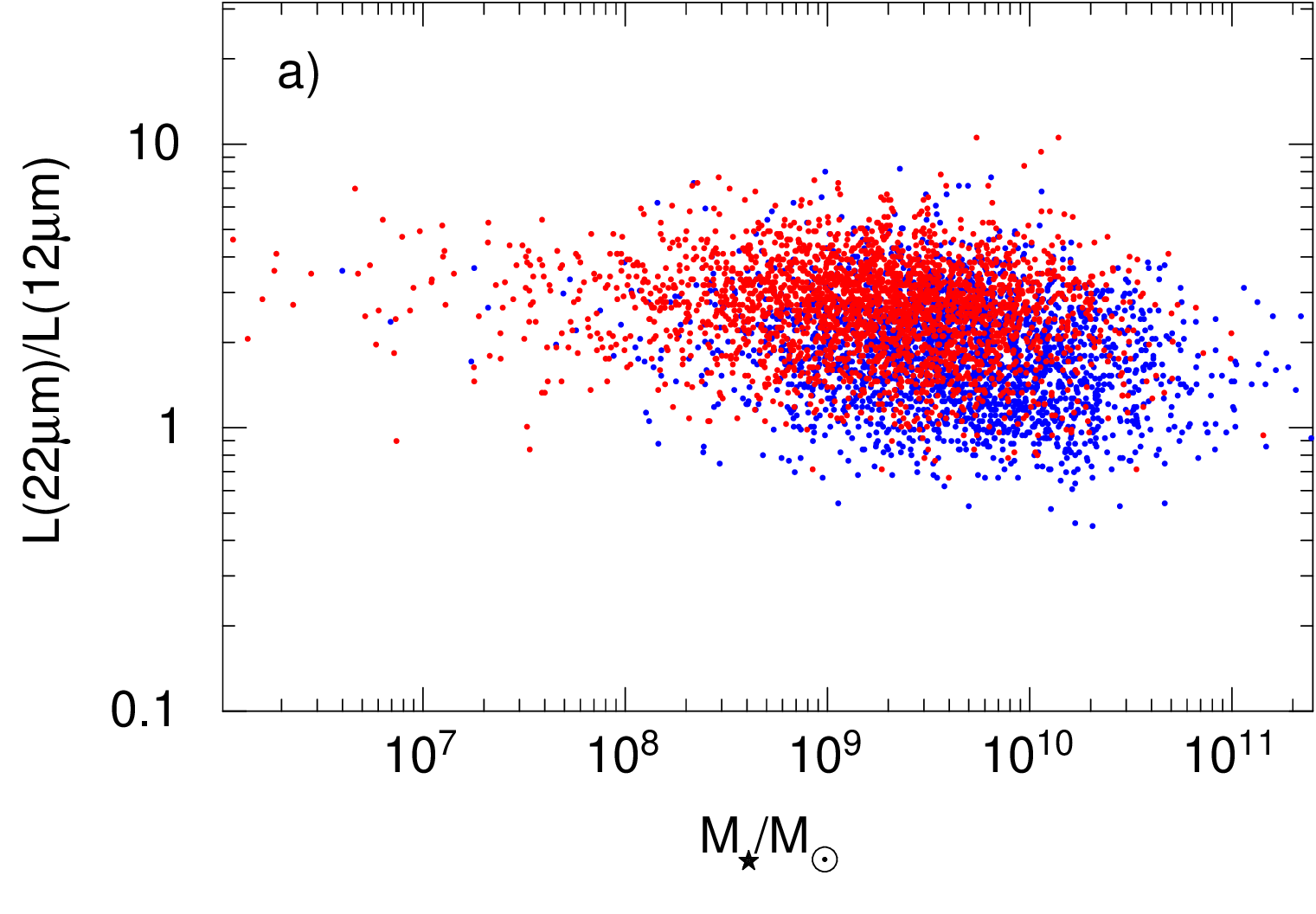}
\includegraphics[angle=-90,width=0.98\linewidth]{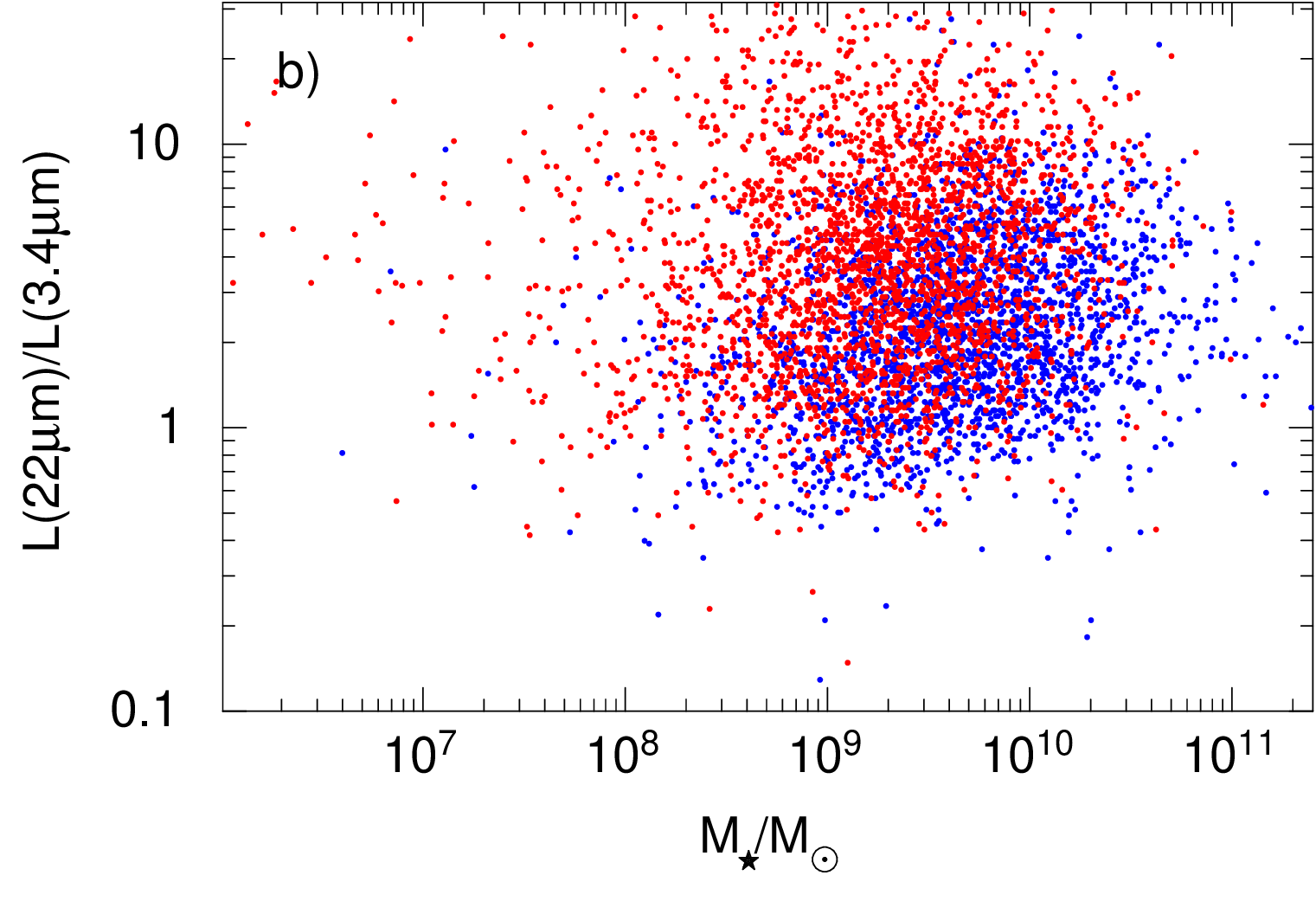}
\caption{{\bf a)} The relation between the 
22$\mu$m-to-12$\mu$m luminosity ratio and the stellar mass.
{\bf b)} The dependence of the
22$\mu$m-to-3.4$\mu$m luminosity ratio on the stellar mass.
CSFGs with EW(H$\beta$) $\geq$ 50\AA\ and EW(H$\beta$) $<$ 50\AA\ 
are shown by red and blue symbols, respectively}
\label{fig4}
\end{figure}

\begin{figure}
\includegraphics[angle=-90,width=0.95\linewidth]{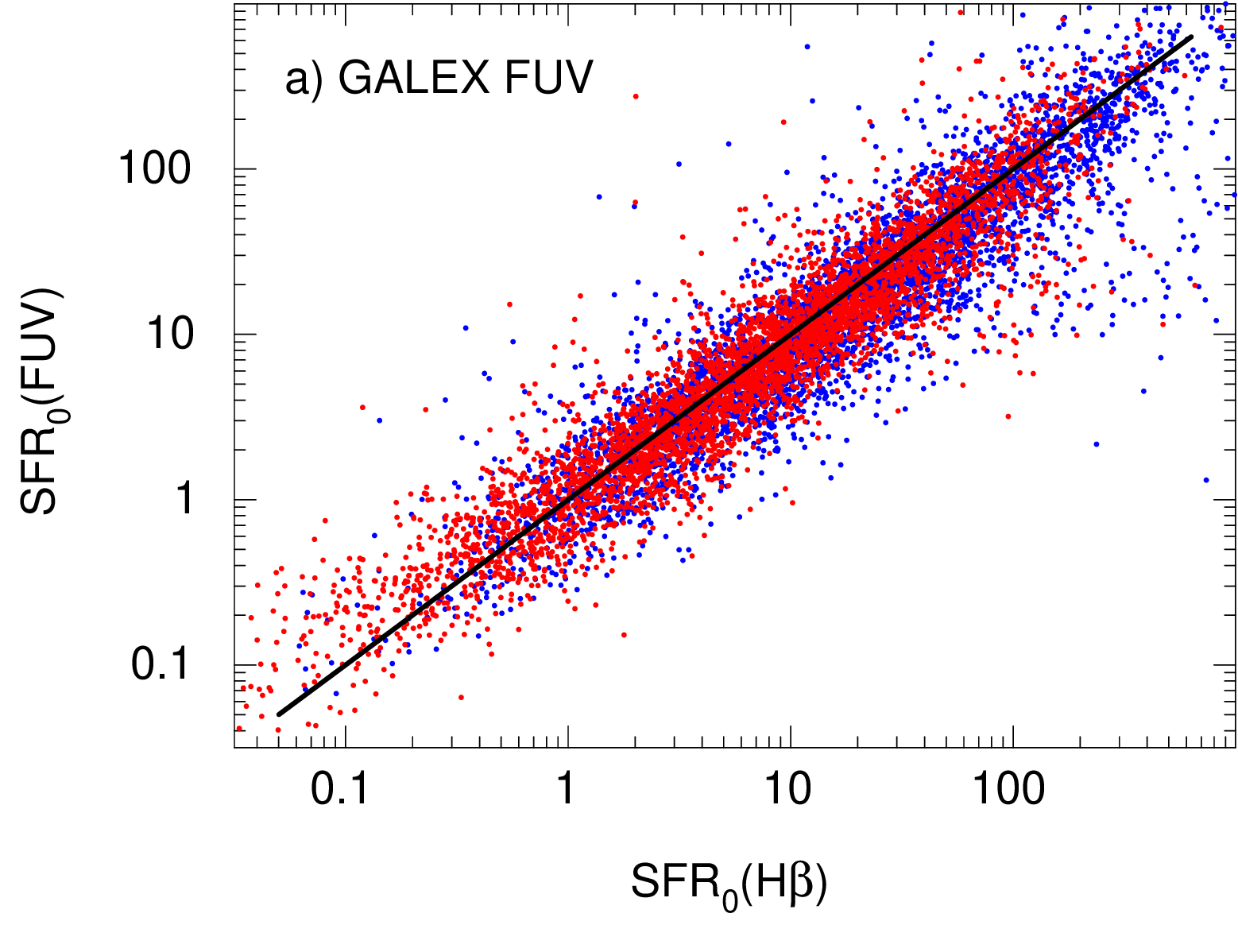}
\includegraphics[angle=-90,width=0.95\linewidth]{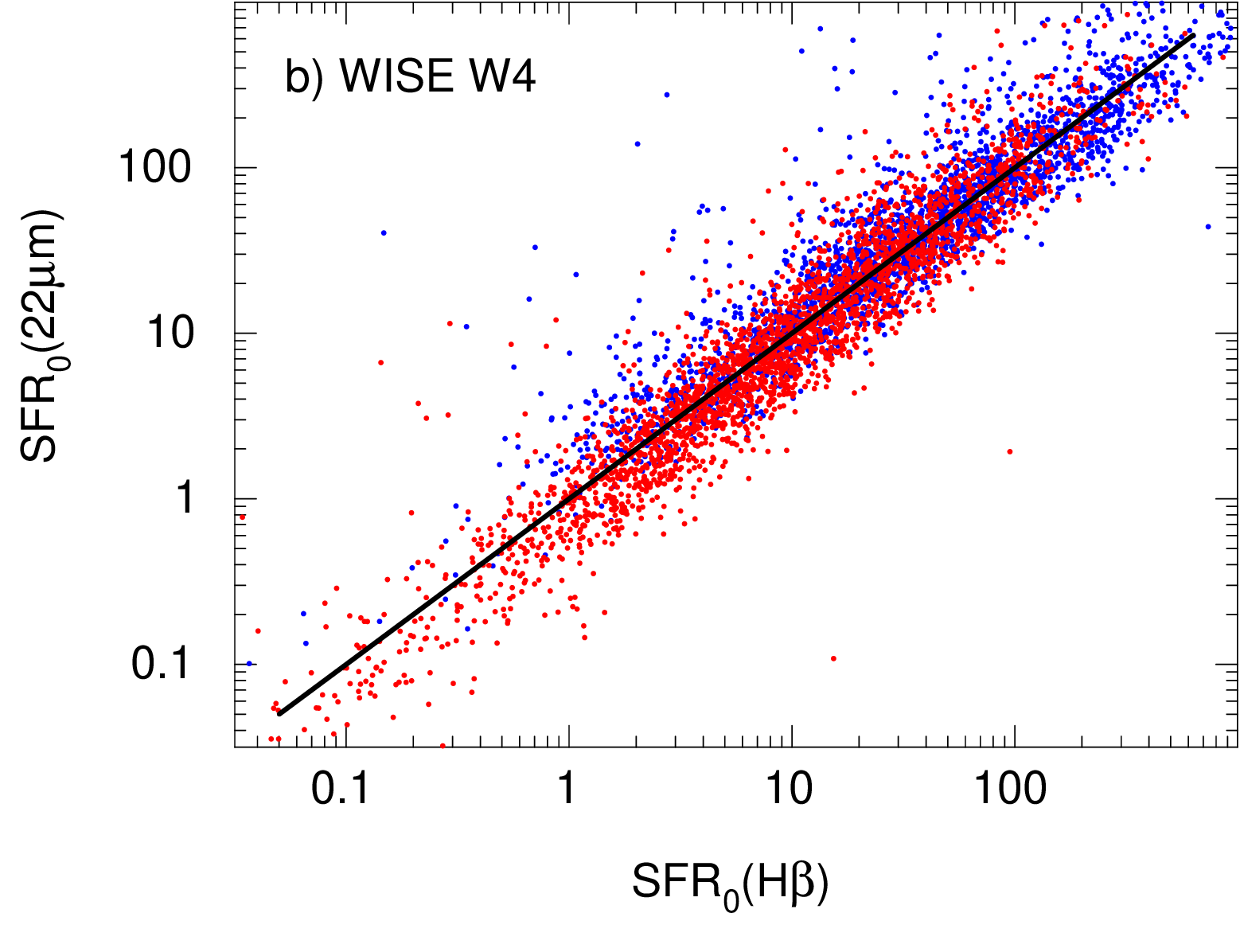}
\includegraphics[angle=-90,width=0.95\linewidth]{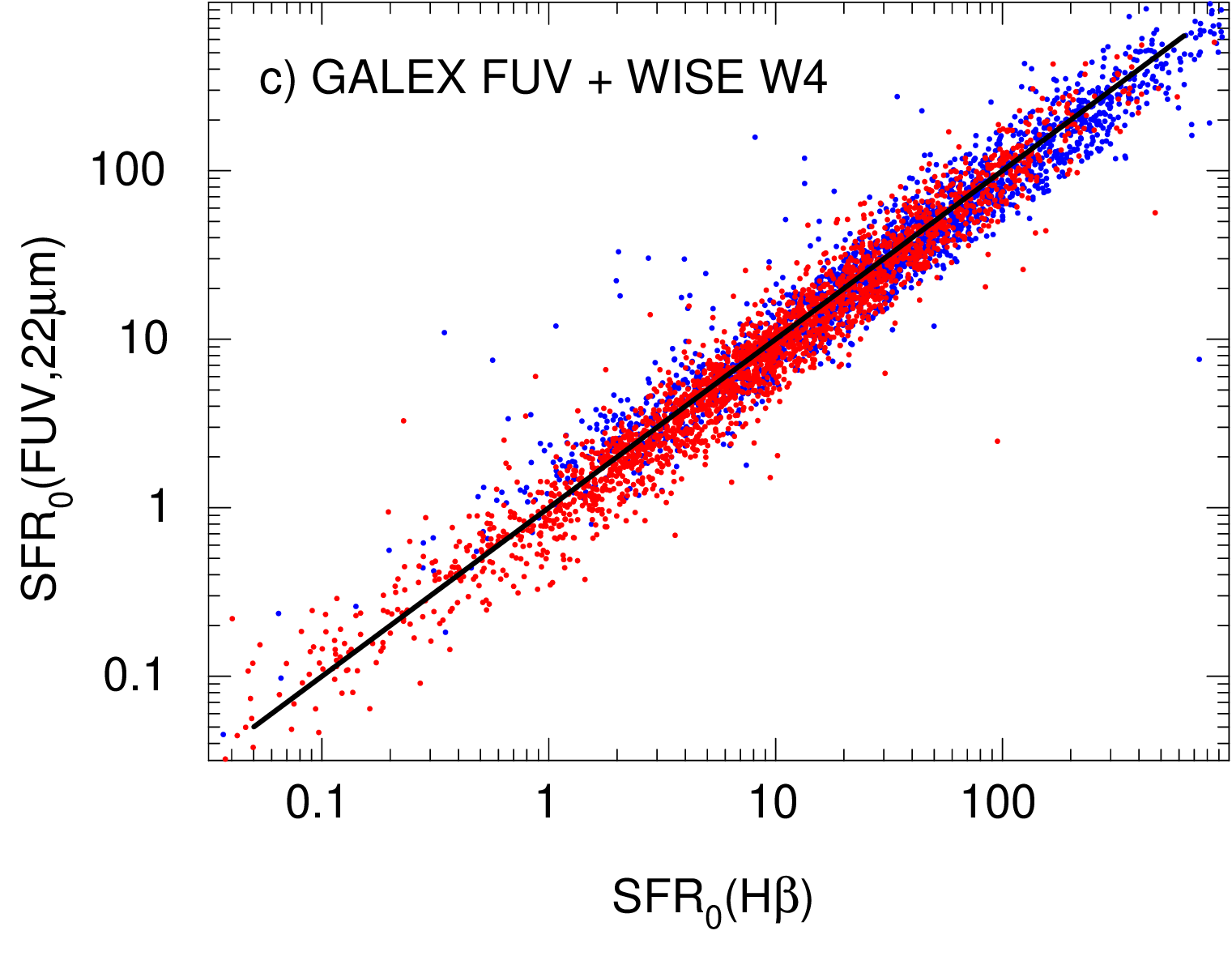}
\caption{Relations between the star formation rate SFR$_0$(H$\beta$)
(Eq. \ref{SFR0Hb}) and {\bf a)} the star formation rate SFR$_0$(FUV)
(Eq. \ref{SFR0FUV}), {\bf b)} the star formation rate SFR$_0$(22$\mu$m) 
(Eq. \ref{SFR0W4}) and {\bf c)} the star formation rate 
SFR$_0$(FUV, 22$\mu$m) (Eq. 7).
All SFRs are expressed in M$_\odot$~yr$^{-1}$. They are derived from
luminosities corrected for the starburst age. 
CSFGs with EW(H$\beta$) $\geq$ 50\AA\ and EW(H$\beta$) $<$ 50\AA\ 
are shown by red and blue symbols, respectively.
Solid lines in all panels are one by one relations}
\label{fig5}
\end{figure}

In this paper, according to \citet{I16a} we adopt that youngest stellar
population in CSFGs with an age of a few Myr is formed
in instantaneous bursts. In this case both the H$\beta$ luminosity, which
is produced by ionizing radiation of the most-massive stars,
and the EW(H$\beta$) rapidly decrease by more than two orders of magnitude
on a time scale of 10 Myr \citep[Fig.~\ref{fig2}a,b; ][]{L99}. We note that 
there is a 
dependence of relations in Fig. \ref{fig2} on the metallicity. 
Similarly, the FUV luminosity also decreases, but more
slowly, because it is produced by less massive and thus more long-lived
stars. \citet{I16a} proposed the following correction to reduce the 
observed H$\beta$ luminosity to the luminosity at a zero starburst age for a 
metallicity of 1/5 solar (dashed line in Fig.~\ref{fig2}c) corresponding to 
typical metallicities of CSFGs from our sample:

\begin{equation}
\Delta\log L({\rm H}\beta) = 2.7 - \log [{\rm EW}({\rm H}\beta)],
\label{corrHb}
\end{equation}
for log EW(H$\beta$) $\leq$ 2.7, otherwise $\Delta$log$L$(H$\beta$) = 0.
Similarly,
\begin{equation}
\Delta\log L({\rm FUV}) = 0.39 \times \Delta\log L({\rm H}\beta).
\label{corrFUV}
\end{equation}

However, we do not apply corrections Eqs. \ref{corrHb} and \ref{corrFUV} 
to the data for CSFGs. Instead we use cubic spline fits of the relation
in Fig.~\ref{fig2}c for the H$\beta$ luminosity and corresponding fits for
the FUV luminosity (not shown) to apply luminosity corrections for the 
respective metallicity of the galaxy. We find that there is no difference 
between the FUV-to-H$\beta$ luminosity ratios reduced to a zero starburst age 
for the galaxies with high observed EW(H$\beta$)s 
(red symbols in Fig. \ref{fig1}b) and the
galaxies with low observed EW(H$\beta$)s 
(blue symbols in 
Fig. \ref{fig1}b). The dispersion of data in Fig. \ref{fig1}b is also 
considerably reduced as compared to that in Fig. \ref{fig1}a.

In Fig. \ref{fig3}a for our sample we show the relation of the 
{\sl WISE} 22$\mu$m-to-H$\beta$ luminosity ratio on the 
aperture- and extinction-corrected H$\beta$ luminosity.
The presence of a tight correlation between the 22$\mu$m and
H$\beta$ luminosities \citep[Fig. \ref{fig3}a, ][]{I14} 
allows us to use the infrared luminosity for the
determination of SFR despite the fact that only a fraction of the UV radiation
is absorbed by dust and re-emitted in the IR range.
It is seen that this relation is 
nonlinear because the $L$(22$\mu$m)/$L$(H$\beta$) ratio increases with 
$L$(H$\beta$). This finding is in accord with many previous studies
\citep[e.g. ][]{C10,I14} and may be attributed to the presence of warmer 
dust in galaxies with the higher $L$(H$\beta$), 
in agreement with predictions of \citet{DL07}.

At variance with FUV-to-H$\beta$ luminosity ratios shown in Fig. \ref{fig1}a,
there is no difference between the galaxies with high and low EW(H$\beta$)s
in Fig. \ref{fig3}a.
Since 22$\mu$m emission is produced by dust, the relation in Fig. \ref{fig3}a
implies that dust in CSFGs is mainly heated by the same stellar population
which produces H$\beta$ emission, in accord with
conclusion made by \citet{I14}. This conclusion is further supported
by the absence of any dependence of the $L$(22$\mu$m)/$L$(H$\beta$) ratio
on extinction in $A(V)$, shown in Fig. \ref{fig3}b. In addition, the 
distribution in Fig. \ref{fig3}b suggests that there is no need to correct
$L$(22$\mu$m) for reddening, particularly due to low extinction in the 
optical range.

The absence of an offset between distributions of CSFGs with low and high 
EW(H$\beta$) (Fig. \ref{fig3}a) implies that reduction of
the 22$\mu$m luminosity to a zero starburst age should be done in the same way 
as that for the H$\beta$ luminosity, i.e. using Eq.~\ref{corrHb}. This is in 
contrast with the results of many other papers where it has been concluded
that infrared emission in normal 
star-forming galaxies is produced by a population continuously formed over 
a period $\ga$ 100 Myr \citep[e.g. ][]{C10}.

We note that the {\sl WISE} 12$\mu$m luminosity can also be used as a SFR 
indicator.
The $L$(22$\mu$m)/$L$(12$\mu$m) ratio is shown in Fig. \ref{fig4}a as a function
of the stellar mass of the galaxy $M_\star$. It is seen that this ratio
is fairly constant with small dispersion of individual CSFGs, indicating
a tight correlation between the two luminosities. The 12$\mu$m band at 
variance with the 22$\mu$m band includes emission of polycyclic aromatic 
hydrocarbons (PAH) which can be heated by both star-forming regions
and older stellar populations. However, \citet{Wu06} have shown
that PAH emission is absent in the most metal-poor blue compact dwarf (BCD) 
galaxies and its strength is generally suppressed 
in a low-metallicity environment. Our CSFGs in this respect are similar
to BCDs by having high-excitation H {\sc ii} regions and low-metallicity.
Furthermore, \citet{H10} showed that the fraction of PAH
emission normalized to the total infrared (IR) luminosity is considerably 
smaller in metal-poor BCDs ($\sim$0.5\%) than that in metal-rich 
SFGs ($\sim$10\%).

Contrary to that the {\sl WISE} emission at 3.4$\mu$m is dominated by the 
light of
cool stars and does not correlate with 22$\mu$m (Fig. \ref{fig4}b). Therefore,
3.4$\mu$m luminosity is not good SFR indicator.

\begin{figure*}
\hbox{
\includegraphics[angle=-90,width=0.49\linewidth]{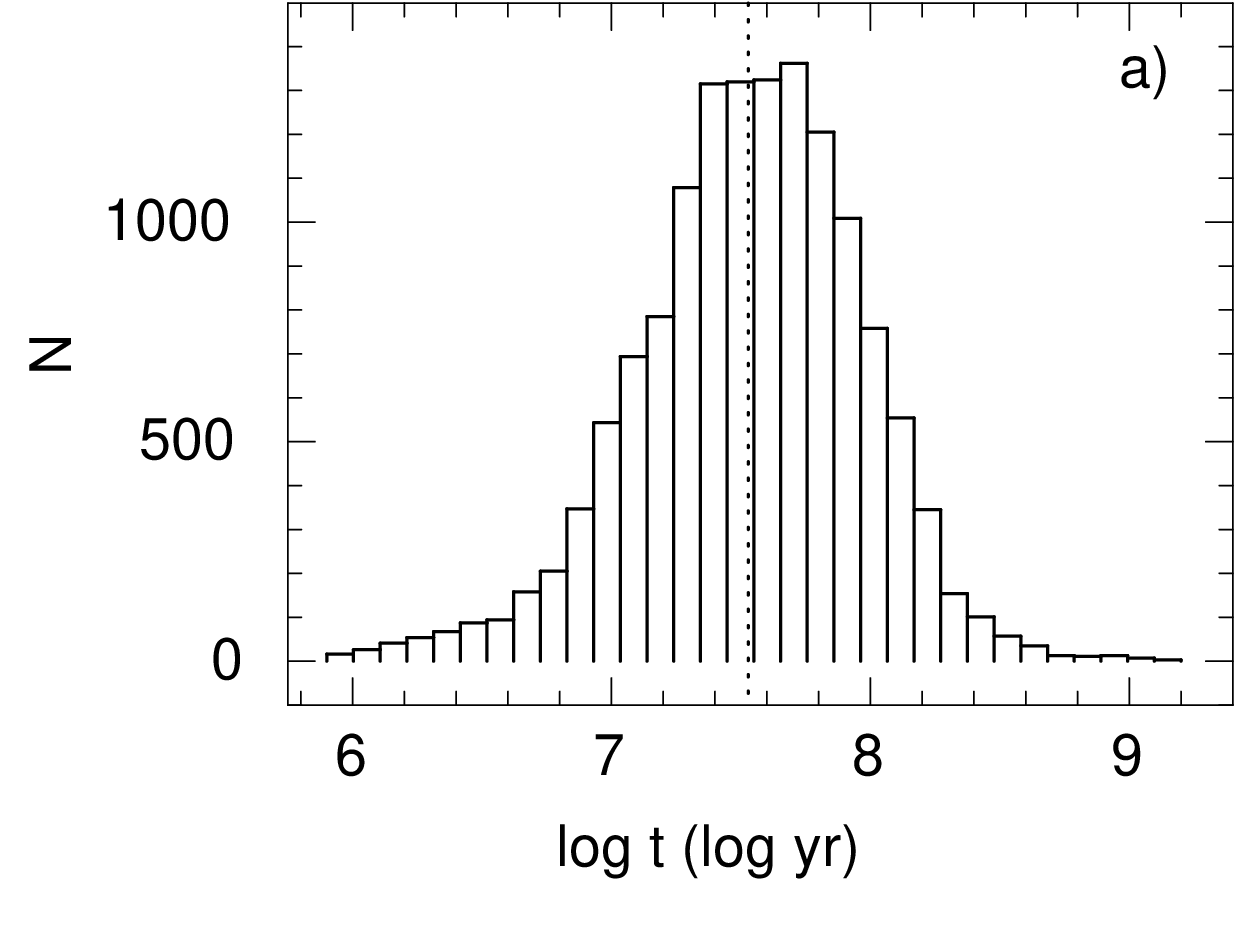}
\includegraphics[angle=-90,width=0.49\linewidth]{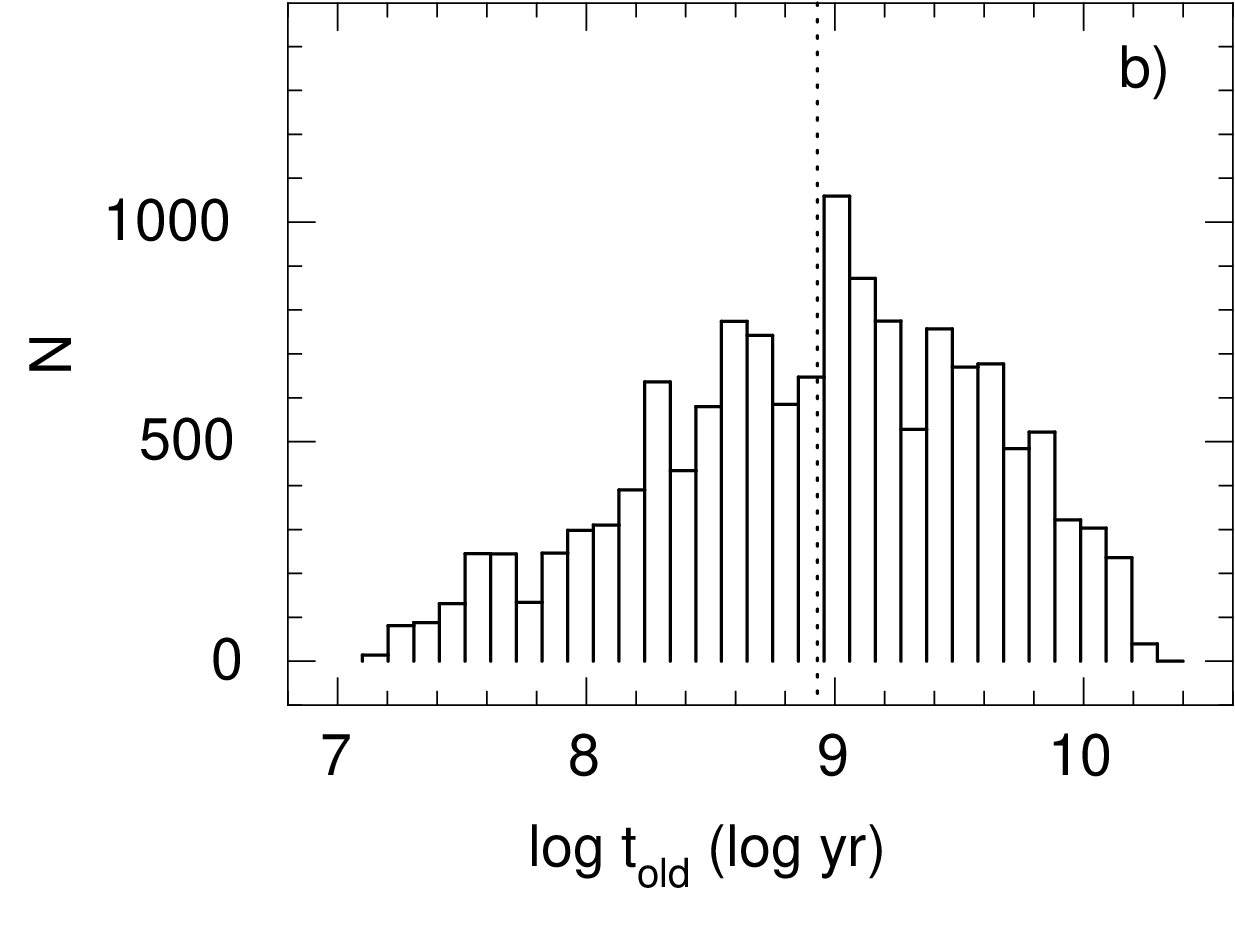}
}
\hbox{
\includegraphics[angle=-90,width=0.49\linewidth]{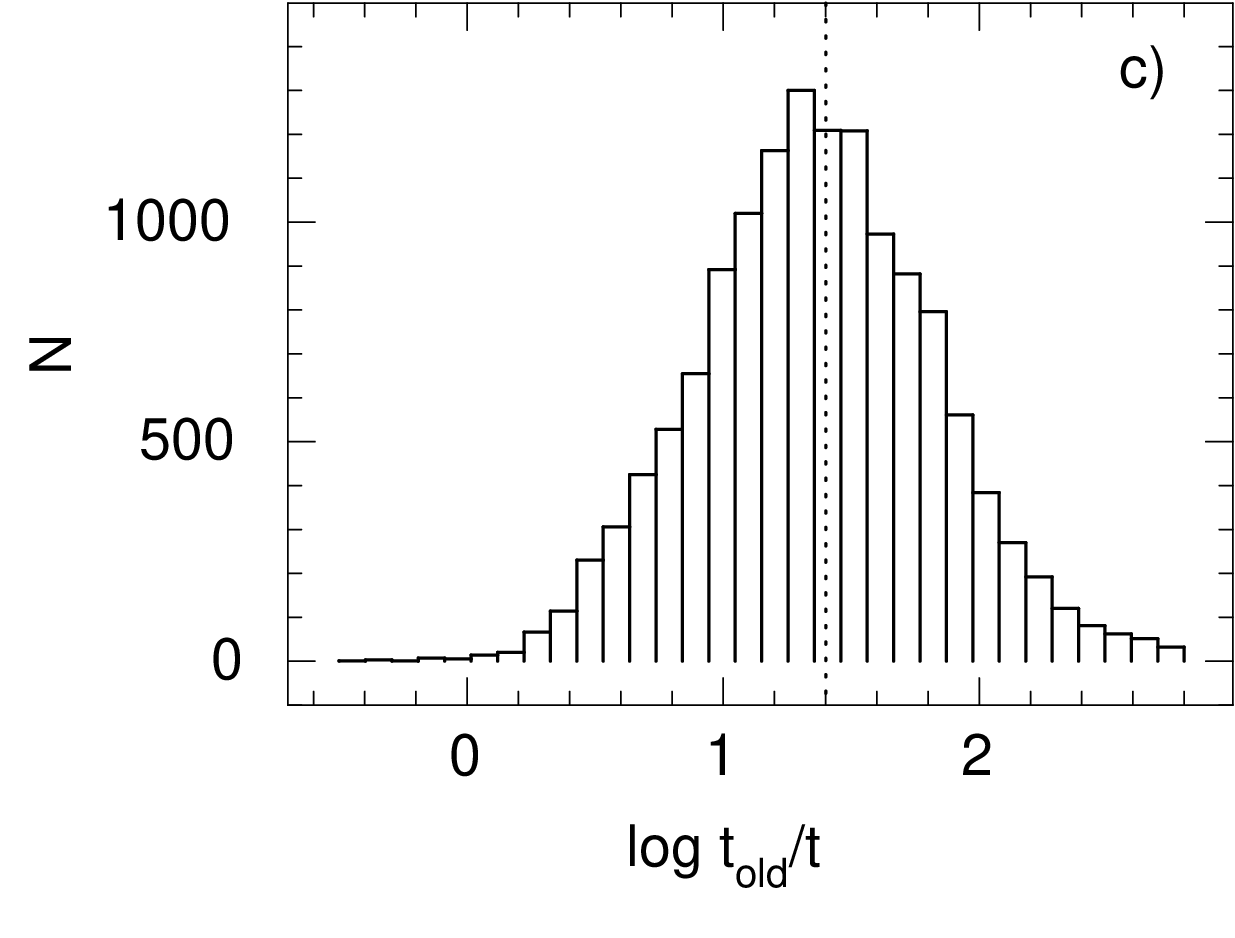}
\includegraphics[angle=-90,width=0.49\linewidth]{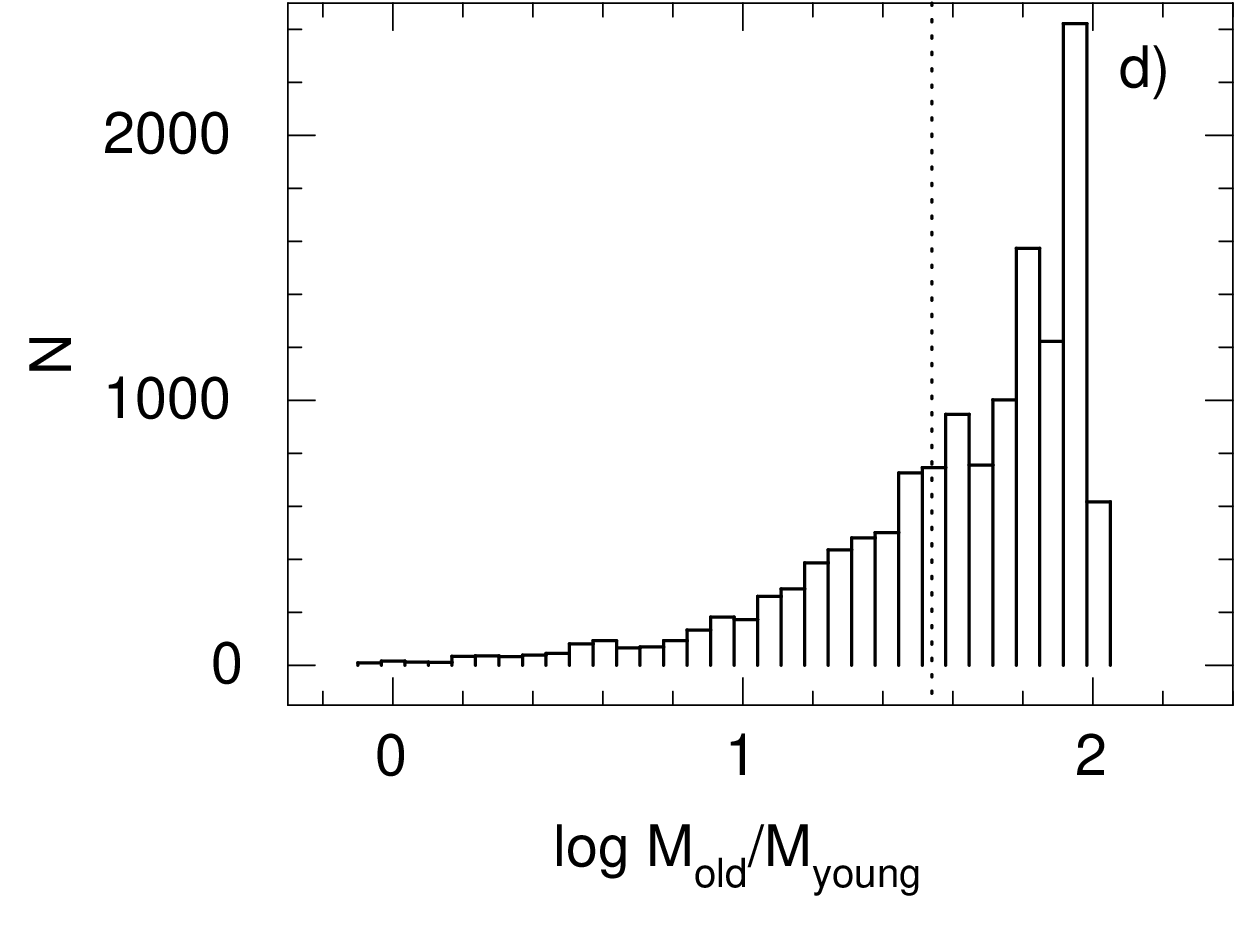}
}
\caption{Histograms of {\bf a)} $t$ = $M_\star$/SFR$_0$(H$\beta$),
{\bf b)} the age of the oldest stars $t_{\rm old}$ derived from SED fitting,
{\bf c)} $t_{\rm old}$/$t$, and {\bf d)} the mass ratio $M_{\rm old}$/$M_{\rm young}$ 
of the old stellar population with the age $>$ 10 Myr 
continuously formed with a constant SFR to the young stellar population formed 
in the burst of star formation with the age $<$ 10 Myr. $M_\star$, $M_{\rm young}$,
$M_{\rm old}$ and $t_{\rm old}$ are derived from SED fitting \citep{I16a}}
\label{fig6}
\end{figure*}

\section{Star formation rates with luminosities reduced to zero starburst age}
\label{SFR0}  

One of the aims of this paper is to find consistent relations for SFRs derived 
from emission in the UV, H$\beta$ and infrared ranges for the sample of
CSFGs. To achieve this we first 
reduce all luminosities to a zero starburst age by using equations in 
Sect.~\ref{SFH}.
Second, we derive the star formation rate SFR$_0$(H$\beta$) from the H$\beta$
luminosity at a zero starburst age adopting 
$L$(H$\alpha$)~=~2.8$\times$$L$(H$\beta$) and using the relation of \citet{K98}:
\begin{equation}
\log {\rm SFR}_0({\rm H}\beta) = \log L_0({\rm H}\beta) - 40.655,
\label{SFR0Hb}
\end{equation}
where log $L_0$(H$\beta$) = log $L$(H$\beta$) + $\Delta$ log $L$(H$\beta$),
SFR$_0$ is in M$_\odot$ yr$^{-1}$ and luminosities are in erg s$^{-1}$.
The SFR$_0$(H$\beta$) is linked to the mass of the young stellar 
population $M_{\rm young}$, formed in the burst, by a simple relation
\begin{equation}
{\rm SFR}_0({\rm H}\beta) \sim 2.8\times 10^{-7} \frac{M_{\rm young}}{M_\odot}. \label{SFR0Hb1}
\end{equation}
Here we use the relation Eq. \ref{SFR0Hb} and the ratio of $\sim$10$^{34.1}$
of the H$\beta$
luminosity at zero age to the stellar mass of a single stellar population by 
\citet{L99} which is only weakly dependent on metallicity.

Third, we adopt that SFR$_0$(H$\beta$), SFR$_0$(FUV), and SFR$_0$(22$\mu$m)
are equal at a zero starburst age. From these assumptions and
using relations between luminosities at respective wavelengths
\citep{I14}, we find
\begin{equation}
\log {\rm SFR}_0({\rm FUV}) = 0.84 \log L_0({\rm FUV}) -  36.02,
\label{SFR0FUV}
\end{equation}

\begin{equation}
\log {\rm SFR}_0(22{\mu}m) = 0.79 \log L_0(22{\mu}m) -  33.50,
\label{SFR0W4}
\end{equation}
where SFR$_0$s are in M$_\odot$ yr$^{-1}$ and luminosities are in erg s$^{-1}$.


Furthermore, we introduce a combination of star-formation
indicators in the FUV and 22$\mu$m ranges, which can be defined as
a geometric mean of SFRs in these two wavelength ranges:
\begin{eqnarray}
\log {\rm SFR}_0({\rm FUV},22{\mu}m)& = 0.5[\log {\rm SFR}_0({\rm FUV}) \\ \nonumber
  & + \log {\rm SFR}_0(22{\mu}m)].
\label{SFR0FW}
\end{eqnarray}

In Fig. \ref{fig5} we show the relations between SFR$_0$'s derived from 
monochromatic and combined SFR indicators. It is seen that
all these relations do not deviate from the one by one 
relations, although the dispersions of points are different. The largest
dispersion is found for the relation between SFR$_0$s derived from the H$\beta$
and FUV luminosities (Fig. \ref{fig5}a), while it is somewhat lower if 
infrared indicators are used (Figs. \ref{fig5}bc). It is also seen that
there are no offsets between the galaxies with low and high observed 
EW(H$\beta$)s (respectively 
blue and red
dots in Fig. \ref{fig5}).
The geometric mean with the lowest dispersion of data is likely the 
most appropriate to use in the case of the CSFG sample (Fig. \ref{fig5}c).

\begin{figure}
\includegraphics[angle=-90,width=0.95\linewidth]{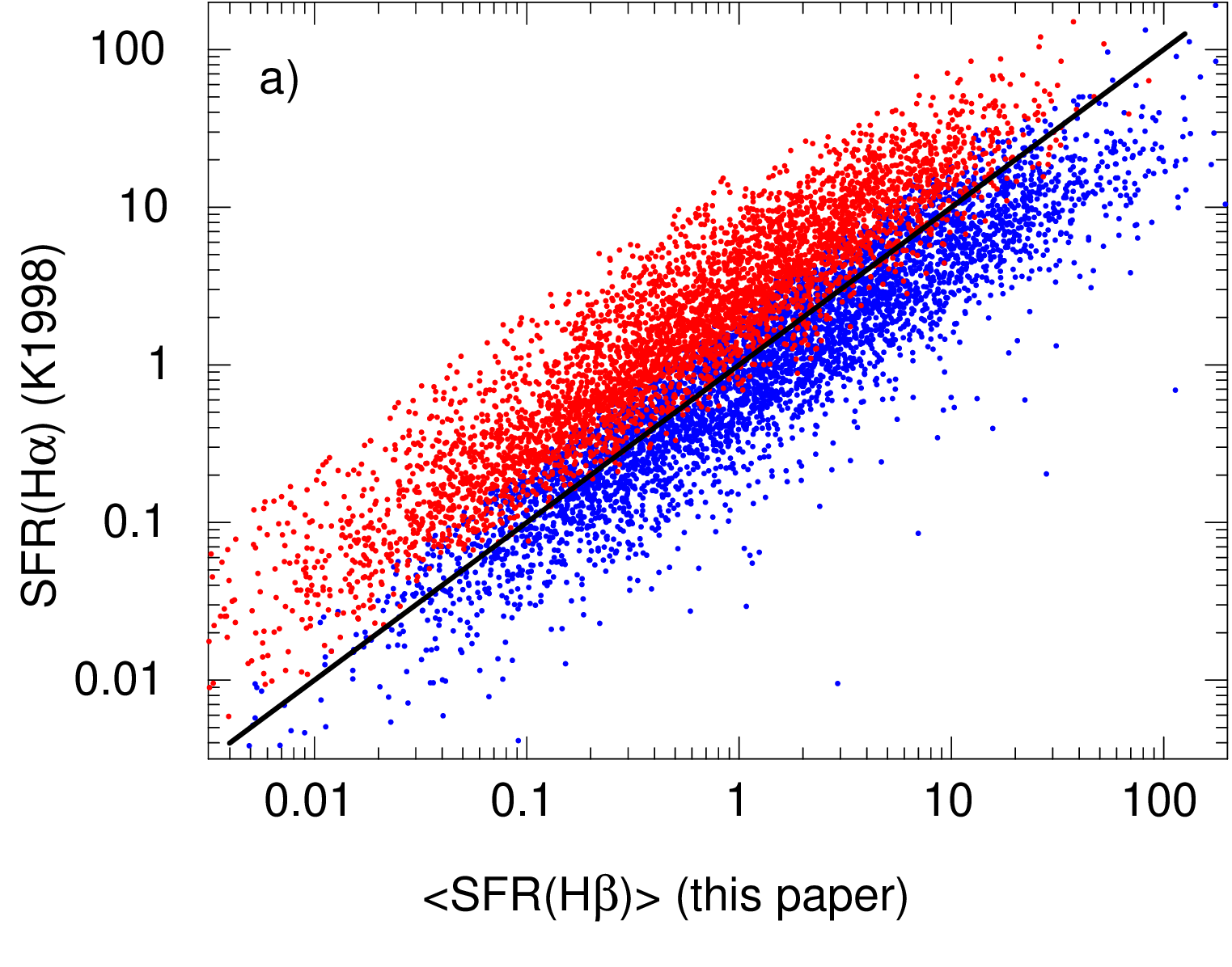}
\includegraphics[angle=-90,width=0.95\linewidth]{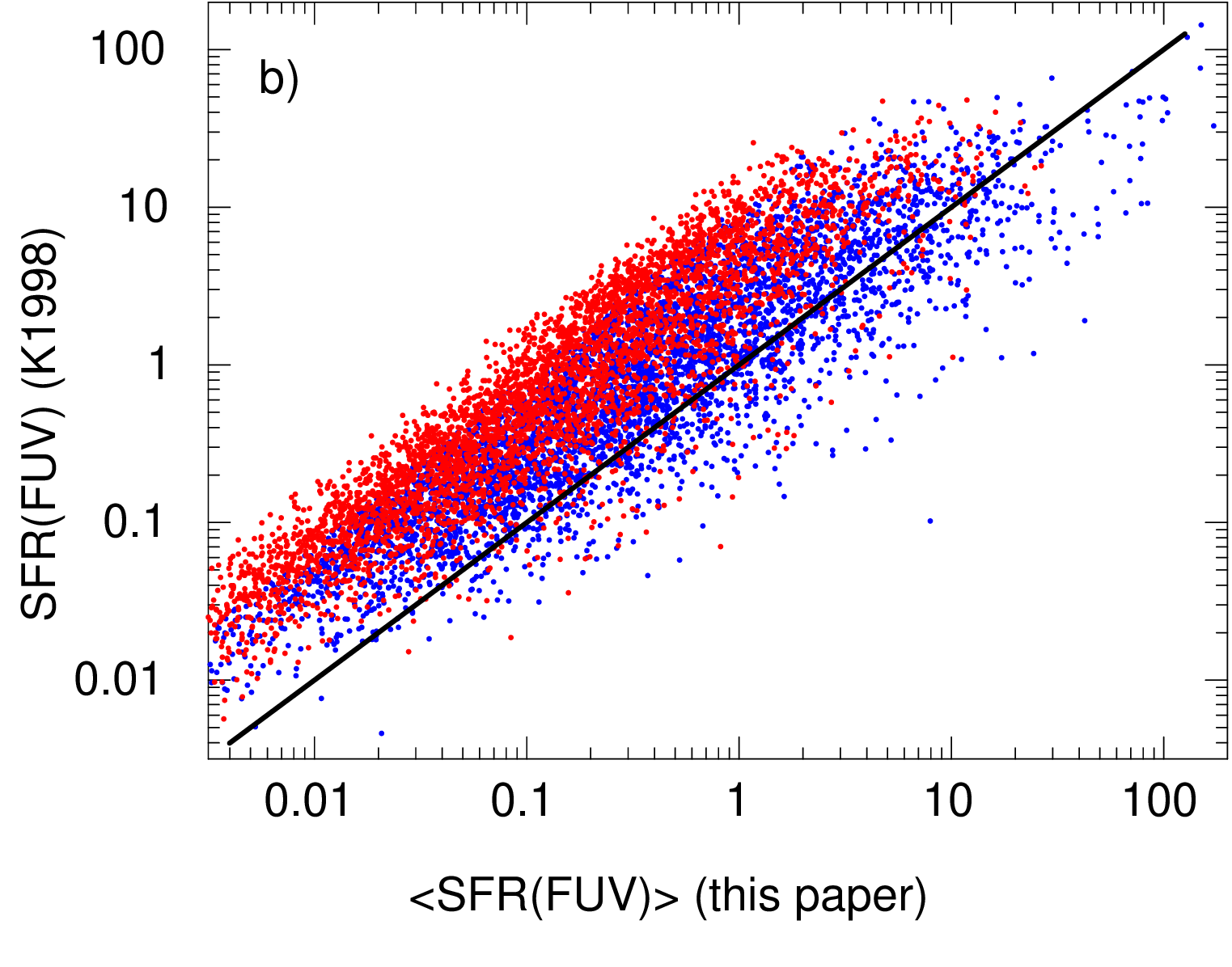}

\includegraphics[angle=-90,width=0.95\linewidth]{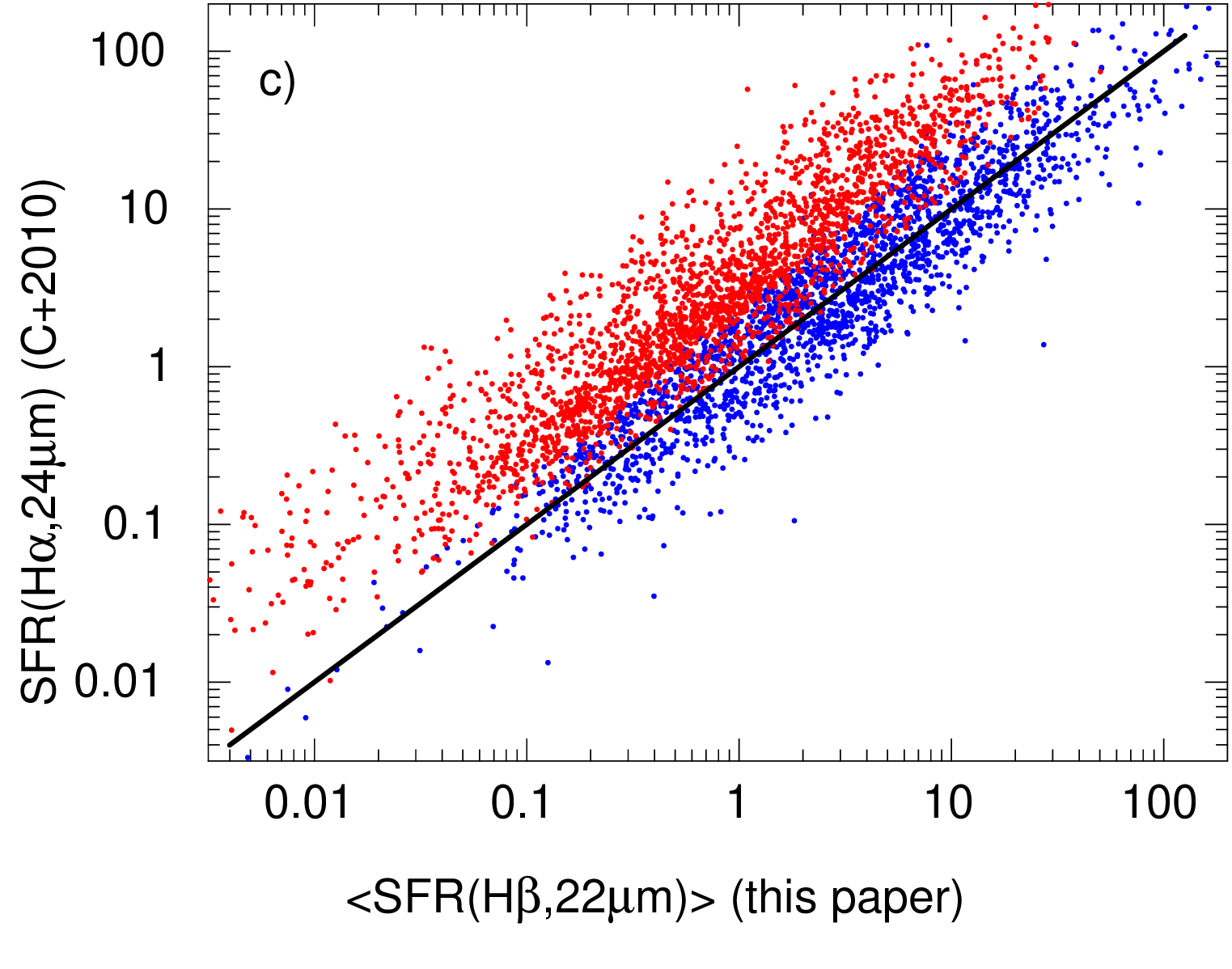}
\caption{The comparison of average star formation rates derived in this
paper and SFRs obtained by other authors. Are shown relations between 
{\bf a)} $<$SFR(H$\beta$)$>$ and SFR(H$\alpha$) by \citet{K98}, {\bf b)} 
$<$SFR(FUV)$>$ and SFR(FUV) by \citet{K98}, and {\bf c)} 
$<$SFR(H$\beta$,22$\mu$m)$>$ and SFR(H$\alpha$,24$\mu$m) by \citet{C10}.
All SFRs are expressed in M$_\odot$ yr$^{-1}$. 
CSFGs with EW(H$\beta$) $\geq$ 50\AA\ and EW(H$\beta$) $<$ 50\AA\ 
are shown by red and blue symbols, respectively.
Solid lines in all panels are one by one relations}
\label{fig7}
\end{figure}

\section{Comparison with star formation rate calibrations by other authors}
\label{SFRcomp}   

In this Section we compare our SFRs with the ones derived by other
authors from the luminosities at the same wavelengths. 
The purpose of this comparison is to understand whether can commonly
used SFR calibrations be applied to derive reasonable SFRs in bursting galaxies
averaged over their lifetime.
Although those authors obtained their relations for SFGs selected
without restrictions on the galaxy morphology, we apply their relations to our
CSFRs. We remind that advantage of our sample is that only relatively
modest corrections for aperture are needed for our compact galaxies when 
observations with different telescopes and different apertures are compared. 
This correction, e.g. of the H$\beta$ flux derived from the SDSS spectrum, for 
extended galaxies can be large and uncertain. Furthermore, the comparison of
SFRs is complicated by the fact that
different assumptions on the star-formation history were made in different
papers, most commonly the continuous star formation over a some time interval
was assumed. In particular, relations of \citet{K98} are obtained for a period
of $\sim$ 100 Myr for a star formation. We also note that adopting the IMF by 
\citet{K01} instead of the \citet{S55} one would result in decrease of SFR by 
a factor of 1.44 \citep{K09}.

The determination of the interval of star formation in bursting CSFGs from our 
sample is somewhat uncertain.
Therefore, the averaged SFR is strongly dependent
on the value of this interval because of a rapid luminosity decline after the 
instantaneous burst.
Furthermore, the star formation history prior the present starburst should
somehow be taken into account. 

To obtain the averaged SFR we first derive 
the time $t$ = $M_\star$/SFR$_0$ needed to build the mass of the galaxy with
SFR$_0$. The histogram of log $t$ is shown in Fig.~\ref{fig6}a. It is seen
that log $t$ is low and it varies in the relatively narrow range between 7 and 8
for most CSFGs from our sample ($\sim$ 80 percent) with the average 
value of 7.53.

We compare $t$ with the age $t_{\rm old}$ of the oldest stars formed in the 
CSFG. The latter quantity as well as $M_\star$ are derived from SED 
fitting of the SDSS spectra \citep[Sect. \ref{s:Sample}, ][]{I16a}. 
The distribution of log $t_{\rm old}$ is shown in Fig.~\ref{fig6}b.
It is much broader than that of log $t$ in Fig.~\ref{fig6}a due to, in part,
the dominating contribution of the young stellar population to the optical 
continuum in SDSS spectra. This contribution makes the determination of 
$t_{\rm old}$ somewhat uncertain. From Fig.~\ref{fig6}b we derive the average
log $t_{\rm old}$ of 8.93, which is much lower than the age of the Universe, 
indicating that most CSFGs from our sample were formed relatively 
recently, $\la$ 1 -- 2 Gyr ago.

We define the star-formation rate $<$SFR$>$ averaged over the galaxy lifetime
for our bursting CSFGs according to the relation
\begin{equation}
<{\rm SFR}> = {\rm SFR_0}/a, \label{avSFR}
\end{equation}
where $a$ = $t_{\rm old}$/$t$. 

The distribution of log ($t_{\rm old}$/$t$) is shown in Fig.~\ref{fig6}c with
the average value of 1.4, corresponding to $a$ = 25. It is interesting to
compare this distribution with the mass ratio $M_{\rm old}$/$M_{\rm young}$
distribution of the stellar population formed during a recent starburst, and an
older population with the age $>$10 Myr, continuously formed with a constant 
SFR (Fig.~\ref{fig6}d). We find that the average log~($M_{\rm old}$/$M_{\rm young}$)
is 1.54, or is slightly higher than the average log~($t_{\rm old}$/$t$), implying
that, in the case of the bursting star formation, $\sim$ 25 -- 35 starbursts 
with similar strengths are needed during the CSFG lifetime to build up
its stellar mass.

Thus, we find that $<$SFR(H$\beta$)$>$ $\sim$ 0.04$\times$SFR$_0$(H$\beta$). 
Correspondingly, to 
keep SFRs derived from other star-formation indicators in different wavelength
ranges equal to each other, we use the same factor 0.04 to convert respective 
SFR$_0$s at a zero starburst age to SFRs averaged over the galaxy lifetime.






In Fig.~\ref{fig7} we compare the $<$SFR$>$s for CSFGs from our
sample as defined by Eq. \ref{avSFR} with SFRs proposed by \citet{K98}
and \citet{C10} by using various star-formation indicators obtained for the
continuous star formation during
the interval of $\sim$ 100 Myr. Fig.~\ref{fig7}a represents
the relation between the $<$SFR(H$\beta$)$>$ 
and SFR(H$\alpha$) obtained by using the observed H$\alpha$ luminosity and the 
relation
by \citet{K98}. It is seen that SFR derived from the \citet{K98} relation is
$\sim$ 2 times higher than the average SFR derived in this paper. The 
difference is higher for CSFGs with younger starbursts (red symbols) as
evidenced by their high EW(H$\beta$) $\geq$ 50\AA.
Similarly, the $<$SFR(FUV)$>$ in Fig. \ref{fig7}b is $\sim$ 3 times lower than 
the SFR(FUV) obtained from the observed FUV luminosity and using the relation 
by \citet{K98}. 

It was suggested in several papers that
the combination of star formation indicators,
which use both the UV and IR observed luminosities, are more reliable as
compared to those which use a single star formation indicator. To check this,
we compare SFR(H$\alpha$,24$\mu$m) by \citet{C10} with the use of the
observed H$\alpha$
and {\sl Spitzer} 24$\mu$m luminosities, and $<$SFR(H$\beta$,22$\mu$m)$>$ 
as defined by Eqs. 7 and \ref{avSFR}.
Similarly to
Figs. \ref{fig7}a, the SFR with the \citet{C10} relation is $\sim$ 2 times
higher than the $<$SFR$>$ derived in this paper. 
We also note that there is 
an offset between the galaxies with high and low EW(H$\beta$)s. This is because 
the bursting nature of star formation is not taken into account in the 
\citet{K98} and \citet{C10} relations, which use the observed luminosities
instead of the luminosities reduced to a zero starburst age.
On the other hand, we note that SFR calibrations by \citet{K98} and 
\citet{C10}
can be used for approximate estimates within a factor of $\sim$ 2 
of the SFR in bursting galaxies averaged
over their lifetime. These estimates would be useful for the determination
of global cosmic chemical enrichment and the intensity of intergalactic 
radiation.


\section{Conclusions} \label{concl}   

In this paper we discuss the relations for the determination of the star
formation rate (SFR) in a large sample of $\sim$ 14000 bursting
compact star-forming galaxies (CSFGs) 
selected by \citet{I16a} from the Data Release 12 (DR12) of the Sloan Digital 
Sky Survey (SDSS).
These data are supplemented by the {\sl GALEX} and {\sl WISE} photometric data
in the far-ultraviolet (FUV) and mid-infrared (MIR) ranges, respectively. We 
argue that the bursting nature of 
star formation in these galaxies casts doubts about applicability of commonly 
used relations for the determination of SFR \citep[e.g. ][]{K98,C10}. This
difference in the star-formation history is taken into account in this paper. 
Our main results are as follows.

1. We obtain relations for the determination of SFR$_0$s at a zero starburst
age from the 
aperture- and extinction-corrected H$\beta$ luminosities, extinction-corrected 
FUV luminosities, and observed 22$\mu$m 
luminosities, which are all reduced to a zero starburst age. We find 
the tightest relation between SFR$_0$(H$\beta$) and 
a geometric mean of star formation rates in the FUV and at 22$\mu$m,
SFR$_0$(FUV,22$\mu$m), while the relations between SFR$_0$(H$\beta$) and
SFR$_0$(FUV), and SFR$_0$(H$\beta$) and SFR$_0$(22$\mu$m) have considerably 
higher dispersions. We attribute these differences to statistical
uncertainties in FUV and 22$\mu$m luminosities.

2. Using SFR$_0$s and results of SED fitting of the SDSS spectra we search for
the recipe to estimate
the star formation rate $<$SFR$>$ averaged over the lifetime of the galaxy. 
We propose to use the relation $<$SFR$>$ = SFR$_0$/$a$ adopting
$a$ = $t_{\rm old}$/$t$, where $t_{\rm old}$ is the age of the oldest stars and
$t$ is a time needed to build a stellar mass of the galaxy $M_\star$ with a 
constant star formation rate SFR$_0$. The average value of $a$ is 25 and
therefore $<$SFR$>$ = 0.04$\times$SFR$_0$.

3. We compared relations between $<$SFR$>$s and commonly used SFRs by 
\citet{K98} and \citet{C10} for our sample of galaxies and found that former
ones are $\sim$ 2 -- 3 lower.
Our relations for the SFR determination are likely more appropriate for CSFGs 
because they take into account rapid temporal evolution of their luminosities.
However, the commonly used SFR calibrations can also be applied for 
approximate estimates within a factor about 2 of the SFR averaged over the 
lifetime of the bursting galaxy.

\section*{Acknowledgments}
We thank the referee whose useful comments helped to make the paper 
clearer. SDSS-III  has  been  provided  by  the  Alfred  P.  Sloan  Foundation,
the Participating Institutions, the National Science
Foundation, and the U.S. Department of Energy Office of
Science. The  SDSS-III  web  site  is  http://www.sdss3.org/.
SDSS-III is managed by the Astrophysical Research Consortium for the 
Participating Institutions of the SDSS-III  Collaboration. GALEX is a NASA 
mission managed by the Jet Propulsion Laboratory. 
This publication makes use of data products from the Wide-field Infrared
Survey Explorer, which is a joint project of the University of California, Los
Angeles, and the Jet Propulsion Laboratory, California Institute of Technology,
funded by the National Aeronautics and Space Administration.

\label{lastpage}


\begin{thebibliography}{}

\bibitem[\protect\citeauthoryear{Alam et al.}{2015}]{A15} Alam, S. et al. 
2015, \apjs, 219, 12 

\bibitem[\protect\citeauthoryear{Aller}{1984}]{A84} Aller,  L.  H.  1984, 
Physics  of  Thermal  Gaseous  Nebulae (Dordrecht: Reidel)


\bibitem[\protect\citeauthoryear{Battisti et al.}{2015}]{B15} Battisti, A. J., 
Calzetti, D., Johnson, B. D., Elbaz D. 2015, \apj, 800, id.143 

\bibitem[\protect\citeauthoryear{Boquien et al.}{2016}]{B16} Boquien, M.
et al. 2016, \aap, 591, A6 

\bibitem[\protect\citeauthoryear{Calzetti et al.}{2010}]{C10} Calzetti, D.
et al. 2010, \apj, 714, 1256 

\bibitem[\protect\citeauthoryear{Cardelli et al.}{1989}]{C89} Cardelli, J. A.,
Clayton, G. C., Mathis, J. S. 1989, \apj, 345, 245


\bibitem[\protect\citeauthoryear{Curtis-Lake et al.}{2013}]{C13} Curtis-Lake, E.
et al. 2013, \mnras, 429, 302 

\bibitem[\protect\citeauthoryear{Draine \& Li}{2007}]{DL07} Draine, B. T.,
Li, A. 2007, \apj, 657, 810

\bibitem[\protect\citeauthoryear{Duarte Puertas et al.}{2017}]{D17} 
Duarte Puertas, S. et al. 2017, \aap, 599, A71 

\bibitem[\protect\citeauthoryear{Girardi et al.}{2000}]{G00} Girardi, L., 
Bressan, A., Bertelli, G., Chiosi, C. 2000, \aaps, 141, 371

\bibitem[\protect\citeauthoryear{Hao et al.}{2011}]{H11} Hao, C.-N.
et al. 2011, \apj, 741, 124 

\bibitem[\protect\citeauthoryear{Hunt et al.}{2010}]{H10} Hunt, L. K., Thuan,
T. X., Izotov, Y. I., Sauvage, M. 2010, \apj, 712, 164

\bibitem[\protect\citeauthoryear{Izotov et al.}{1994}]{I94} 
Izotov, Y. I., Thuan, T. X., Lipovetsky, V. A. 1994, \apj, 435, 647

\bibitem[\protect\citeauthoryear{Izotov et al.}{2014}]{I14} 
Izotov, Y. I., Guseva, N. G., Fricke, K. J., Henkel, C. 2014, \aap, 561, A33


\bibitem[\protect\citeauthoryear{Izotov et al.}{2016a}]{I16a} 
Izotov, Y. I., Guseva, N. G., Fricke, K. J., Henkel, C. 2016a, \mnras, 462, 
4427

\bibitem[\protect\citeauthoryear{Izotov et al.}{2016b}]{I16b} 
Izotov, Y. I. et al. 2016b, \mnras, 461, 3683

\bibitem[\protect\citeauthoryear{Izotov et al.}{2018}]{I18} 
Izotov, Y. I. et al. 2018, \mnras, 474, 4514

\bibitem[\protect\citeauthoryear{Jiang et al.}{2013}]{J13} 
Jiang, L. et al. 2013, \apj, 772, id.99 

\bibitem[\protect\citeauthoryear{Karman et al.}{2017}]{K17} 
Karman, V. et al. 2017, \aap, 599, A28 

\bibitem[\protect\citeauthoryear{Kashino et al.}{2013}]{K13} Kashino, D. et al.
2013, \apj, 777, L8 

\bibitem[\protect\citeauthoryear{Kennicutt}{1998}]{K98} Kennicutt, R. C. Jr.
1998, \araa, 36, 189

\bibitem[\protect\citeauthoryear{Kennicutt et al.}{2009}]{K09} 
Kennicutt, R. C. Jr. et al. 2009, \apj, 703, 1672

\bibitem[\protect\citeauthoryear{Kroupa}{2001}]{K01} Kroupa, P.
2001, \mnras, 322, 231

\bibitem[\protect\citeauthoryear{Lee et al.}{2013}]{L13} 
Lee, J. C., Hwang, H. S., Ko, J. 2013, \apj, 774, id. 62 

\bibitem[\protect\citeauthoryear{Leitherer et al.}{1999}]{L99}
Leitherer, C. et al. 1999, \apjs, 123, 3

\bibitem[\protect\citeauthoryear{Leitherer et al.}{2014}]{L14} Leitherer, C.
et al.
2014, \apjs, 212, 14

\bibitem[\protect\citeauthoryear{Lejeune et al.}{1997}]{L97} 
Lejeune, T., Buser, R., Cuisinier, F. 1997, \aaps, 125, 229


\bibitem[\protect\citeauthoryear{Panella et al.}{2015}]{P15}
Panella, M. et al. 2015, \apj, 807, id. 141 

\bibitem[\protect\citeauthoryear{Planck Collaboration XVI}{2014}]{P14} 
Planck Collaboration XVI 2014, \aap, 571, A16

\bibitem[\protect\citeauthoryear{Puglisi et al.}{2016}]{P16}
Puglisi, A. et al. 2016, \aap, 583, id. A83 

\bibitem[\protect\citeauthoryear{Reddy et al.}{2010}]{R10}
Reddy, N. A. et al. 2010, \apj, 712, 1070 

\bibitem[\protect\citeauthoryear{Reddy et al.}{2015}]{R15}
Reddy, N. A. et al. 2015, \apj, 806, id. 259 

\bibitem[\protect\citeauthoryear{Salpeter}{1955}]{S55} Salpeter, E. E. 1955, 
\apj, 121, 161

\bibitem[\protect\citeauthoryear{Schmutz et al.}{1992}]{S92} Schmutz, W., 
Leitherer, C., Gruenwald, R. 1992, \pasp, 104, 1164

\bibitem[\protect\citeauthoryear{Shivaei et al.}{2015}]{S15}
Shivaei, I., Reddy, N. A., Steidel, C. C., Shapley, A. E. 2015, \apj, 804, id.149 

\bibitem[\protect\citeauthoryear{Shivaei et al.}{2016}]{S16}
Shivaei, I. et al. 2016, \apj, 820, L23 

\bibitem[\protect\citeauthoryear{Weisz et al.}{2016}]{W11} Weisz, D. R. 2011, 
\apj, 739, 5

\bibitem[\protect\citeauthoryear{Wright}{2006}]{W06} Wright, E. L. 2006, 
\pasp, 118, 1711

\bibitem[\protect\citeauthoryear{Wu et al.}{2006}]{Wu06} Wu, Y., Charmandaris,
V., Hao, L., Brandl, B. R., Bernard-Salas, J., Spoon, H. W. W., Houck, J. R. 
2006, \apj, 639, 157

\bibitem[\protect\citeauthoryear{Yang et al.}{2017}]{Y17} 
Yang, H. et al. 2017, \apj, 844, 171

\bibitem[\protect\citeauthoryear{Zhang et al.}{2013}]{Z13} 
Zhang, F., Li, L., Kang, X., Zhuang, Y., Han, Z. 2013, \mnras, 433, 1039 


\end{thebibliography}
\end{document}